\documentclass[aps,twocolumn,prd,superscriptaddress,nofootinbib]{revtex4-1}

\usepackage{mathtools}
\usepackage{amsfonts}
\usepackage{mathrsfs}
\usepackage{bbm}
\usepackage{slashed}
\usepackage{amsmath}
\usepackage{bm}

\usepackage{graphicx}
\usepackage{color}
\usepackage{colortbl}
\usepackage{array}

\usepackage{float}
\usepackage{placeins}
\usepackage{booktabs}
\usepackage[caption=false]{subfig}
\usepackage{makecell}
\usepackage{tabstackengine}

\usepackage{xspace}
\usepackage{siunitx}
\usepackage{hyperref}
\usepackage[nameinlink]{cleveref}
\usepackage{bookmark}

\usepackage{xifthen}
\usepackage{xcolor}
\hypersetup{
	colorlinks,
	linkcolor={red!75!black},
	citecolor={blue!75!black},
	urlcolor={blue!75!black}
}
\usepackage{units}

\usepackage[utf8]{inputenc}

\def\sec#1{Sec.~\ref{#1}}

\def\eq#1{\labelcref{#1}}

\def\s#1{{\scriptscriptstyle #1}}

\setkeys{Gin}{width=0.48\textwidth}
\newcommand{\tr}{{\text{tr}}}

\newcommand*{\eg}{e.g.\@\xspace}
\newcommand*{\ie}{i.e.\@\xspace}

\newcommand{\imag}{\text{i}}

\newcommand{\gettitle}{Ghost spectral function from the spectral Dyson-Schwinger equation}

\newcommand{\getHeidelbergAffiliation}{\affiliation{Institut f\"ur Theoretische Physik, Universit\"at Heidelberg, Philosophenweg 16, 69120 Heidelberg, Germany}}

\newcommand{\getEMMIAffiliation}{\affiliation{ExtreMe Matter Institute EMMI, GSI, Planckstr. 1, 64291 Darmstadt, Germany}}

\hypersetup{
	colorlinks,
	linkcolor={red!75!black},
	citecolor={blue!75!black},
	urlcolor={blue!75!black},
	pdftitle={\gettitle},
	pdfauthor={Horak,Pawlowski, Wink},
	pdfkeywords={analytic continuation}
	{correlations functions} {scalar theory}
	{functional renormalisation group}
	{real time} {spectral function} 
	bookmarksopen=true,
	bookmarksopenlevel=2,
	bookmarksnumbered=true
}

\allowdisplaybreaks

\begin{document}

\title{\gettitle}

\author{Jan Horak}
\getHeidelbergAffiliation

\author{Joannis Papavassiliou}
\affiliation{Department of Theoretical Physics and IFIC, University of Valencia and CSIC,
E-46100, Valencia, Spain}

\author{Jan M. Pawlowski}
\getHeidelbergAffiliation
\getEMMIAffiliation
  
\author{Nicolas Wink}
\getHeidelbergAffiliation

\begin{abstract}

We compute the  ghost spectral function in Yang-Mills theory by solving the corresponding Dyson-Schwinger equation for a given input gluon spectral function. The results encompass both scaling and decoupling solutions for the gluon propagator input. The resulting ghost spectral function displays a particle peak at vanishing momentum and a negative scattering spectrum, whose infrared and ultraviolet tails are obtained analytically. The ghost dressing function is computed in the entire complex plane, and its salient features are identified and discussed.

\end{abstract}

\maketitle

\section{Introduction} \label{sec:Introduction}

The complete access to the hadronic bound state and resonance structure, as well as to the non-perturbative dynamics of QCD at finite temperature and density, requires the computation of timelike correlation functions. In functional approaches, such as the functional renormalisation group (fRG) and Dyson-Schwinger equations (DSEs), the respective non-perturbative computations are carried out numerically, and hence, demand a numerical approach to timelike correlation functions. 

Recently, the \textit{spectral DSE}  approach has been put forth~\cite{Horak:2020eng}, based on the K\"all\'en-Lehmann (KL) representation of correlation functions in terms of spectral functions. In particular, in~\cite{Horak:2020eng} the general properties of this novel approach, including a consistent spectral renormalisation procedure, have been expounded and applied in the context of a scalar $\phi^4$-theory in 2+1 dimensions. 

As a first step towards a full QCD treatment, we use this spectral DSE for the ghost-gluon system in Yang-Mills theory. In recent years, ghost and gluon spectral functions have been reconstructed from numerical data of Euclidean ghost and gluon propagators, see e.g.\  \cite{Haas:2013hpa, Dudal:2013yva, Cyrol:2018xeq, Binosi:2019ecz, Dudal:2019gvn}. Also direct computations have been put forward, either perturbatively, e.g.\  \cite{Siringo:2016jrc, Hayashi:2020few}, with non-perturbative analytically continued DSEs~\cite{Strauss:2012dg, Fischer:2020xnb}, or in a spectral approach~\cite{Sauli:2020dmx}. While these direct computations unravel highly interesting structures, they are still inconclusive. 

Specifically, in the gluon DSE one has to deal with a rather complicated diagrammatic representation with many non-trivial ingredients, ranging from the presence of one- and two-loop diagrams, the dependence on ghost and gluon propagators, as well as several vertices with complicated momentum dependence. Consequently, some properties of the gluon spectral functions, and in particular the potential presence and location of complex conjugate poles, are rather unstable under small variations of the vertices involved; for a detailed recent discussion, see~\cite{Fischer:2020xnb}. 

\begin{figure*}[t]
	\centering
	\includegraphics[width=0.85\linewidth]{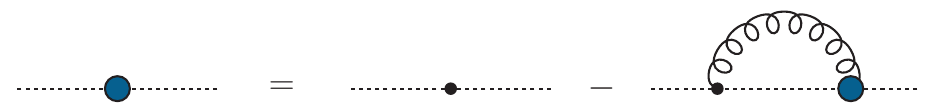}
	\caption{Diagrammatic representation of the Dyson-Schwinger equation for of the inverse ghost propagator. Blue dots represent full 1-PI vertices. Internal lines stand for full propagators.
	}
	\label{fig:ghost_DSE}
\end{figure*}

Instead, the ghost DSE, see \Cref{fig:ghost_DSE}, requires far less non-trivial input: apart from the ghost propagator, it depends only on the gluon propagator or, equivalently, the gluon spectral function, as well as the ghost-gluon vertex. The latter is protected by non-renormalisation, and hence shows a very mild momentum-dependence. Accordingly, in the present work we approximate this vertex by its classical counterpart. 

This leaves us with a rather stable set-up: the spectral ghost DSE is solved on the basis of given input gluon spectral functions, obtained by appropriately modifying the result of~\cite{Cyrol:2018xeq}, which was reconstructed under the assumption of a KL representation of the gluon. We also test the stability of the result under a variation of the input by tuning the whole family of scaling and decoupling solutions.   

This work is organised as follows: In \Cref{sec:YM-KL} we discuss spectral properties of Yang-Mills theory. In~\Cref{sec:ghost_DSE}, the spectral ghost DSE is set up, and the input gluon spectral function is discussed. We present our results for the ghost propagator and spectral function in~\Cref{sec:results}, and discuss our findings in~\Cref{sec:Conclusion}. The appendix includes some technical details.

\section{Yang-Mills theory and the spectral representation}
\label{sec:YM-KL}

We consider $3+1$-dimensional Yang-Mills theory with three colours, $N_c=3$, in the Landau gauge. The gauge-fixed classical action including the ghost action reads, 
\begin{align} \label{eq:YM_action}
	S_{\text{YM}} = \int_x \frac{1}{4} \tr F_{\mu\nu} F_{\mu\nu} + \frac{1}{2\xi} (\partial_\mu A_\mu^a)^2 - \bar{c}^a \partial_\mu D_\mu^{ab} c^b \,,
\end{align}
where $\xi$ denotes the gauge fixing parameter and \mbox{$\int_x = \int d^4 x$}. The Landau gauge is achieved for $\xi = 0$. Note that in \labelcref{eq:YM_action} we have chosen a positive dispersion for the ghost. The field strength, $F_{\mu\nu}$, and covariant derivative, $D_\mu$, in the adjoint representation are given by
\begin{align} \label{eq:field_strength_cov_derivative} \nonumber
	F_{\mu\nu}^a = & \, \partial_\mu A_\nu^a - \partial_\nu A_\mu^a + g f^{abc} A_\mu^b A_\nu^c \,, \\[1ex]
	D_\mu^{ab} = & \, \delta^{ab} \partial_\mu - g f^{abc} A_\mu^c \,, 
\end{align}
with the structure constants $f^{abc}$ of SU(3). 
\begin{figure}[b]
	\centering
	\subfloat{\includegraphics[width=0.22\textwidth]{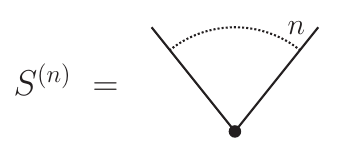}}\hspace{3mm}
	\subfloat{\includegraphics[width=0.22\textwidth]{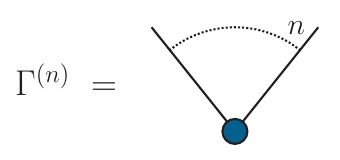}} \\
	\subfloat{\includegraphics[width=0.49\textwidth]{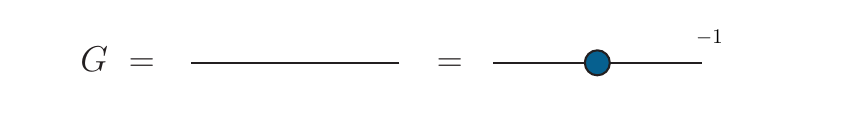}}
	\caption{Diagrammatic notation used throughout this work: Lines stand for full propagators, small black dots stand for classical vertices, and larger blue dots stand for full vertices.}
	\label{fig:FunctionalMethods:DSE_notation}
\end{figure}

Functional relations such as the flow equations in the fRG or the DSEs are one- and two-loop exact relations for the full correlation functions. For reviews see~\cite{Pawlowski:2005xe, Gies:2006wv, Rosten:2010vm, Braun:2011pp, Pawlowski:2014aha, Dupuis:2020fhh} (fRG) and~\cite{Roberts:1994dr, Alkofer:2000wg, Maris:2003vk, Fischer:2006ub, Binosi:2009qm, Maas:2011se,  Huber:2018ned} (DSEs). The pivotal r$\hat{\textrm{o}}$le in these diagrammatic relations is played by the connected two-point functions $\langle \phi \phi\rangle_\textrm{c}$, the propagators of the theory. They read 
\begin{align}
	\langle \phi_i(p)\phi_j(q)\rangle_\textrm{c} = (2 \pi)^4\delta(p+q) {\cal T}_{\phi_i\phi_j}(p) \, G_{\phi_i}(p)  \,,
\end{align}
where $\phi=(A_\mu, c,\bar c)$, and the tensor $ {\cal T}_{\phi_i\phi_j}(p)$ carries the Lorenz and gauge group tensor structure. The scalar parts of the propagators are given by \mbox{$G_{\phi}=G_A, G_c, -G_c$}. In the Landau gauge, the  gluon propagator is transverse, $[{\cal T}_{A A}(p)]^{ab}_{\mu\nu} =\delta^{ab}\Pi^\bot_{\mu\nu}(p)$, where \mbox{$\Pi_{\mu\nu}^{\perp}(p) = \delta_{\mu\nu} - p_\mu p_\nu / p^2$} denotes the standard transverse projection operator. The scalar part of the gluon propagator reads 
\begin{align}
		G_A(p)=\, \frac{1}{Z_A(p) \, p^2} \,,
	\label{eq:gluon_prop_param}\end{align}
with the gluon dressing function $1/Z_A(p)$. Similarly, for the ghost we have ${\cal T}_{c \bar c}^{ab}= \delta^{ab}$, and the scalar part reads 
\begin{align}
	\label{eq:ghost_prop_param}
 	G_c(p)=\frac{1}{Z_c(p) \, p^2} \,.
\end{align}
with the ghost dressing function $1/Z_c(p)$. In what follows we will compute \labelcref{eq:ghost_prop_param} for time-like momenta. In fact, as we will see below, our results permit the evaluation of $G_c(p)$ for general complex momenta. Extensions of correlation functions to the complex plane are particularly interesting, in view of their relevance for the self-consistent treatment of bound-state problems, see, \eg~\cite{Cloet:2013jya, Eichmann:2016yit, Sanchis-Alepuz:2017jjd}.

If the KL spectral representation~\cite{Kallen:1952zz,Lehmann:1954xi} is applicable, a propagator $G$ can be recast in terms of its spectral function $\rho$, 
\begin{equation} \label{eq:KL}
	G_\phi(p_0,|\vec{p}|) = \int_0^\infty \frac{d\lambda}{\pi}\frac{\lambda\,\rho_\phi(\lambda,|\vec p|)}{p_0^2+\lambda^2} \,.
\end{equation}
The spectral function naturally arises as the set of non-analyticities of the propagator in the complex momentum plane. If \labelcref{eq:KL} holds, the non-analyticities are restricted to the (linear) real momentum axis. Equation \labelcref{eq:KL} leads to the following inverse relation between spectral function and the retarded propagator,
\begin{equation} \label{eq:spec_func_def}
\rho_\phi(\omega,\left|\vec{p}\right|) = 2 \, \text{Im} \, G_\phi(-\text{i} (\omega+\text{i}0^+ ),\left|\vec{p}\right|) \,,
\end{equation} 
where $\omega$ is a real frequency. This formulation allows us to work only with the frequency argument and set the spatial momentum to zero in practice, since the full phase-space can be restored from Lorentz invariance. Hence, for the remainder of this work, $\left|\vec{p}\right|$ will be dropped.

%

Now we detail the above generic spectral representation \labelcref{eq:KL} for the Yang-Mills propagators. We will consider the entire family (decoupling/massive and scaling) of potential infrared solutions, for detailed discussions see \cite{Aguilar:2008xm, Boucaud:2008ky, Fischer:2008uz, Cyrol:2016tym}. In terms of the ghost and gluon propagator inverse dressing functions $Z_c(p)$ and $Z_A(p)$, a generic decoupling behaviour is characterised by
\begin{align} 
	\label{eq:decoupling_param}
	\lim_{p \to 0} Z_A(p) \sim \frac{1}{p^2}\,, \qquad \qquad 
	\lim_{p \to 0} Z_c(p) = Z_c \,, 
\end{align}
with a finite $Z_c=Z_c(p=0)$.

Formally, the ghost propagator is expected to obey the KL-representation~\cite{Lowdon:2017gpp}. Also recent reconstructions show no signs of a violation of this property~\cite{Binosi:2019ecz,Dudal:2019gvn}. The ghost spectral function must exhibit a single particle peak at vanishing spectral value, with residue $1/Z_c$. In addition, a continuous scattering tail is expected to show up in the spectral function via the logarithmic branch cut. This leads us to the general form of the ghost spectral function, 
\begin{align}
	\label{eq:ghost_spec_general}
	\rho_c(\omega) = \frac{\pi}{Z_c} \, \frac{\delta(\omega)}{\omega} + \tilde{\rho}_c(\omega) \,,
\end{align}
where $\tilde{\rho}_c$ denotes the continuous tail of the spectral function and $\delta(\omega)/\omega$ has to be understood as a limiting process $\delta(\omega-m)/\omega$ with $m\to 0^+$. Inserting \labelcref{eq:ghost_spec_general} in \labelcref{eq:KL} leads us to a spectral representation for the dressing function, 
\begin{align} \label{eq:tilderho}
	\frac{1}{Z_c(p)} = \frac{1}{Z_c}  + p^2 \int \frac{d\lambda}{\pi} \frac{\lambda \, \tilde{\rho}_c(\lambda)}{p^2+\lambda^2} \,. 
\end{align}
In the case where the spectral function can be normalised by solely integrating it over the whole branch cut, the normalisation is given by the value of the inverse dressing function at infinity. A detailed derivation of this is given in~\Cref{app:sum_rules}. Since the inverse ghost dressing tends to zero for large momenta, the spectral function obeys
\begin{align} \label{eq:ghost_spec_norm}
	\int \frac{d\lambda}{\pi} \lambda \, \rho_c(\lambda) = 0 \,.
\end{align}
Taking into account the explicit form of $\rho_c$~\labelcref{eq:ghost_spec_general}, we immediately arrive at
\begin{align} \label{eq:spec_norm_uv}
\int \frac{d\lambda}{\pi}\lambda \, \tilde{\rho}_c(\lambda) = - \frac{1}{Z_c} \,. 
\end{align}
Equation \labelcref{eq:ghost_spec_norm}, or its consequence, \labelcref{eq:spec_norm_uv}, are the ghost analogue of the Oehme-Zimmermann super-convergence relation for the gluon, \cite{Oehme:1979ai, Oehme:1990kd}; for a generic discussion, see~\Cref{app:sum_rules} and~\cite{Bonanno:2021squ}. 

The present work is based on the assumption that the gluon propagator admits a KL representation of the form given in \labelcref{eq:KL}, with $\rho_\phi\to \rho_A$. Note that the validity of this assumption is subject of an ongoing debate; for results and discussions, see, e.g.~\cite{Dudal:2008sp, Sorella:2010it, Strauss:2012dg, Haas:2013hpa, Lowdon:2017uqe, Cyrol:2018xeq, Hayashi:2018giz,  Lowdon:2018uzf, Binosi:2019ecz, Li:2019hyv, Fischer:2020xnb, Hayashi:2021nnj, Hayashi:2020few, Kondo:2019ywt,Kondo:2019rpa}. We will employ the reconstruction results of~\cite{Cyrol:2018xeq} for the gluon spectral function $\rho_A$, based on the Euclidean propagator data of~\cite{Cyrol:2016tym}.
\begin{figure}[t]
	\centering
	\includegraphics[width=\linewidth]{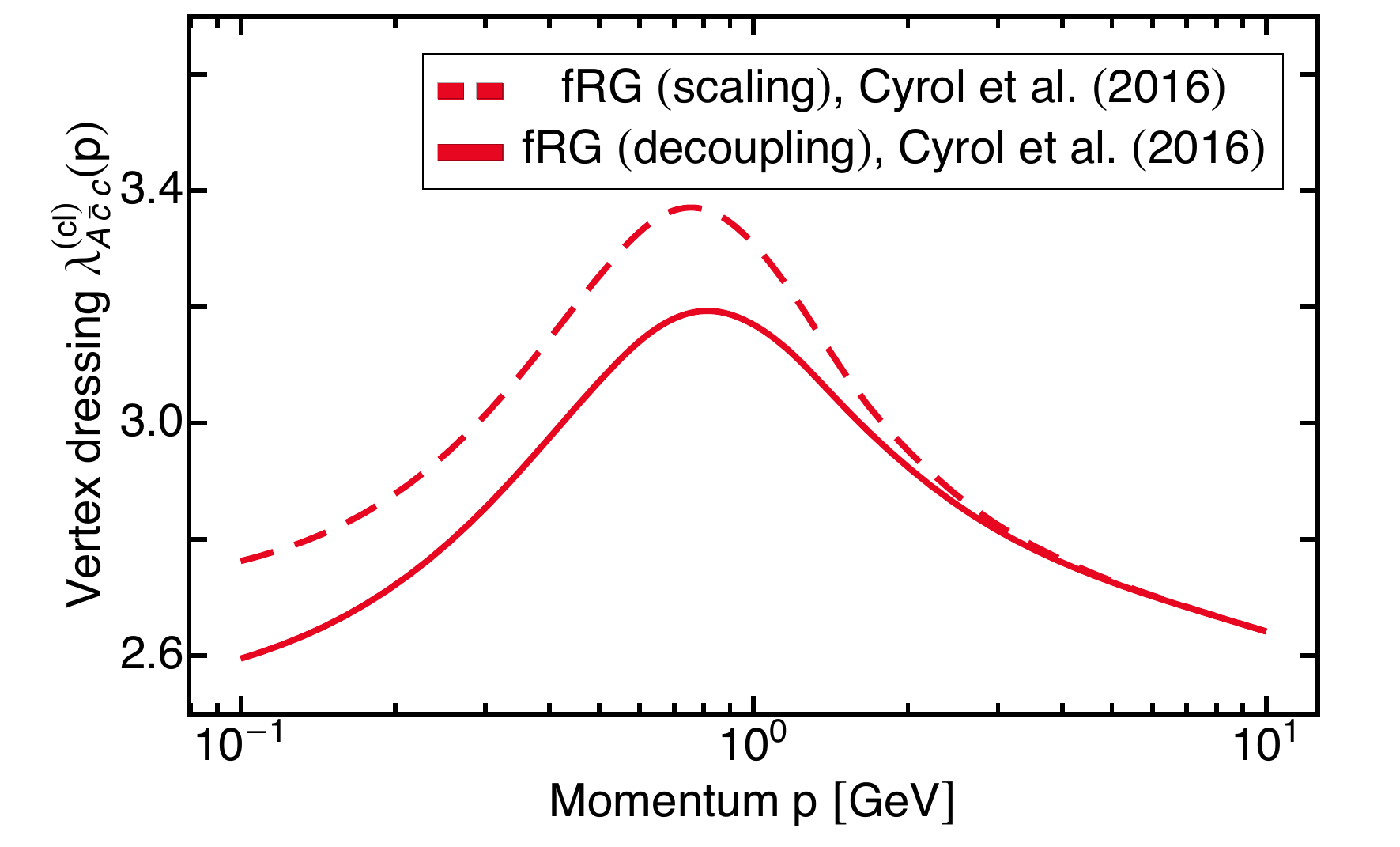}
	\caption{Ghost-gluon vertex dressing $\lambda^{(\textrm{cl})}_{A c\bar c}(p,q)$, see \labelcref{eq:ggl-vertex}, data taken from~\cite{Cyrol:2018xeq}. The dressing is shown at the symmetric point $p^2=q^2=(p+q)^2$ for scaling and lattice-type decoupling solution, more details can be found in~\cite{Cyrol:2018xeq}.}
	\label{fig:ghost_gluon_vert_FRG}
\end{figure}
%

\section{The spectral ghost DSE in Yang-Mills Theory}
\label{sec:ghost_DSE}

In this section, we use the preparations in \Cref{sec:YM-KL} to set up the spectral DSE for the ghost spectral function $\rho_c$. The DSE for the ghost propagator in Yang-Mills theory is depicted in~\Cref{fig:ghost_DSE}, while the diagrammatic notation employed is summarised in~\Cref{fig:FunctionalMethods:DSE_notation}. As discussed before, its only input is the scalar part $G_A$ of the gluon propagator and the full ghost-gluon vertex. The latter consists of two tensor structures, see e.g.~\cite{Cyrol:2016tym},
\begin{align}\label{eq:ggl-vertex}
	 [\Gamma_{A \bar c c}]^{abc}_\mu(p,q) = \imag f^{abc} \left[ q_\mu \lambda^{ (\textrm{cl}) }_{A \bar c c}(p,q) + p_\mu \lambda^{ (\textrm{nc}) }_{A \bar c c}(p,q)\right]\,,
\end{align}
with incoming gluon momentum $p$ and anti-ghost momentum $q$, and we have dropped the momentum conserving $\delta$-function. 

The ghost-gluon vertex in \labelcref{eq:ggl-vertex} contains two independent vertex dressings, $\lambda^{ (\textrm{cl}) }$ (classical tensor structure) and $\lambda^{ (\textrm{nc})}$ (non-classical). The non-classical dressing is proportional to the gluon momentum and hence drops out of the ghost DSE due to the transversality of the Landau gauge gluon propagator.

The classical dressing is subject to Taylor's non-renormalisation theorem, and has a very mild momentum dependence, see e.g.~\cite{Schleifenbaum:2004id, Sternbeck:2006rd, Ilgenfritz:2006he, Boucaud:2011eh, Dudal:2012zx, Cyrol:2016tym, Aguilar:2013xqa, Huber:2020keu, Barrios:2020ubx}, see \Cref{fig:ghost_gluon_vert_FRG} with the data from~\cite{Cyrol:2016tym}. In \Cref{fig:ghost_gluon_vert_FRG} we depict the dressing $\lambda^{ (\textrm{nc})}$ at the symmetric point $p^2=q^2=(p+q)^2$ for both, the scaling solution as well as the lattice-type decoupling solution. For further explanations we refer to the detailed discussion of~\cite{Cyrol:2016tym}. Accordingly, we consider the approximation 
\begin{align}\label{eq:ghgl-approx}
\lambda^{ (\textrm{cl}) }_{A \bar c c}(p,q)\approx g\,, 
\end{align} 
where $g$ is the gauge coupling at the renormalisation point $\mu_{\s{\textrm{RG}}}$. Within the MOM-type  scheme that we employ, the dressing functions acquire at $\mu_{\s{\textrm{RG}}}$ their tree-level value, \ie, $Z_A(\mu_{\s{\textrm{RG}}})=Z_c(\mu_{\s{\textrm{RG}}})=1$.

We emphasise that our approach is by no means restricted to classical vertices only: quantum corrections may be duly accounted for, as long as the momentum loops involved can by integrated analytically. Especially, upon construction of spectral representations for higher $n$-point-functions, see e.g.,~\cite{Evans:1991ky,Aurenche:1991hi,PhysRevD.49.4107,Guerin_1994,Wink:2020tnu,Carrington_1998,Bros:1996mw,Defu_1998,Weldon_1998,Hou_1998,Hou:1999zv,PhysRevD.72.096005,Bodeker:2017deo}, fully dressed vertices of general form can be included. 

In summary, the ghost gap equation in~\Cref{fig:ghost_DSE} can be written as 
\begin{align}
	\label{eq:ghost_DSE_eucl_general}
		[\Gamma^{(2)}_{\bar c c}]^{ab}(p) = \tilde{Z}_3 \delta^{ab} p^2 - \Sigma_{\bar c c}(p)\delta^{ab} \,,
\end{align}
with the renormalisation constant $\tilde{Z}_3$ associated with the ghost field. Similarly, the classical ghost-gluon vertex in~\Cref{fig:ghost_DSE} contains the respective renormalisation constant, $S_{A c\bar c}^{\mu}(p,q) = - \tilde Z_1 \imag g f^{abc} p^\mu$. The (scalar) ghost self-energy $\Sigma_{\bar c c}(p)$ with the approximation
\labelcref{eq:ghgl-approx} is then given by,   
\begin{align} \label{eq:ghost_self_energy_general}
	\Sigma_{\bar c c}(p) =&\, g^2 C_A \tilde Z_1  \int_q \left( p^2 - \frac{(p\cdot q)^2}{q^2}\right) \,G_A(q) G_c(p+q) \,, 
\end{align}
where $C_A=N_c$ is the eigenvalue of the second Casimir operator of the colour group SU($N_c$) in the adjoint representation.   

The ghost DSE of~\labelcref{eq:ghost_DSE_eucl_general} can be straightforwardly rewritten in terms of the ghost dressing function,  
\begin{align}\label{eq:DSEZc}
	Z_c(p) = \tilde{Z}_3 \delta^{ab} - \frac{\Sigma_{\bar c c}(p)}{p^2} \,.
\end{align}
Now we write \labelcref{eq:DSEZc} in terms of spectral loops by using \labelcref{eq:KL} for ghost and gluon propagators. This leads us to 
\begin{align}\nonumber 
		\Sigma_{\bar c c}(p) = \, & g^2 N_c \int_{\lambda_1, \lambda_2} \lambda_1 \lambda_2 
		\rho_A(\lambda_1) \rho_c(\lambda_2) \\[1ex]
		& \hspace{-1cm} \times \int_q \left( p^2 - \frac{(p\cdot q)^2}{q^2}\right)  \frac{1}{q^2 + \lambda_1^2} 
		\frac{1}{(p+q)^2 + \lambda_2^2} \,,
\label{eq:ghost_self_energy_final}\end{align}
with $\rho_A$ and $\rho_c$ the gluon and ghost spectral functions, respectively. Note that the order of spectral and momentum integration have been interchanged, implicitly assuming the finiteness of the dimensionally regularised expression.

\subsection{Spectral renormalisation}
\label{sec:SpecRenorm}

The momentum integral in \labelcref{eq:ghost_self_energy_final} involves two massive propagators with spectral masses $\lambda_1$ and $\lambda_2$. It is readily evaluated upon introduction of Feynman parameters and using dimensional regularisation in $d=4-2\epsilon$ dimensions. Calculational details and the resulting expression are given in \Cref{app:loop_mom_int}.
\begin{figure*}[t]
	\centering
	\subfloat[Gluon spectral functions (scaling and decoupling), see \labelcref{eq:gluon_spec_dec}, based on the reconstruction of the scaling spectral function in~\cite{Cyrol:2018xeq} (red-dashed line). The spectral functions differ only in the infrared, shown in the inset.  \hspace*{\fill} \label{fig:gluSpecPlot}]{\includegraphics[width=.48\textwidth]{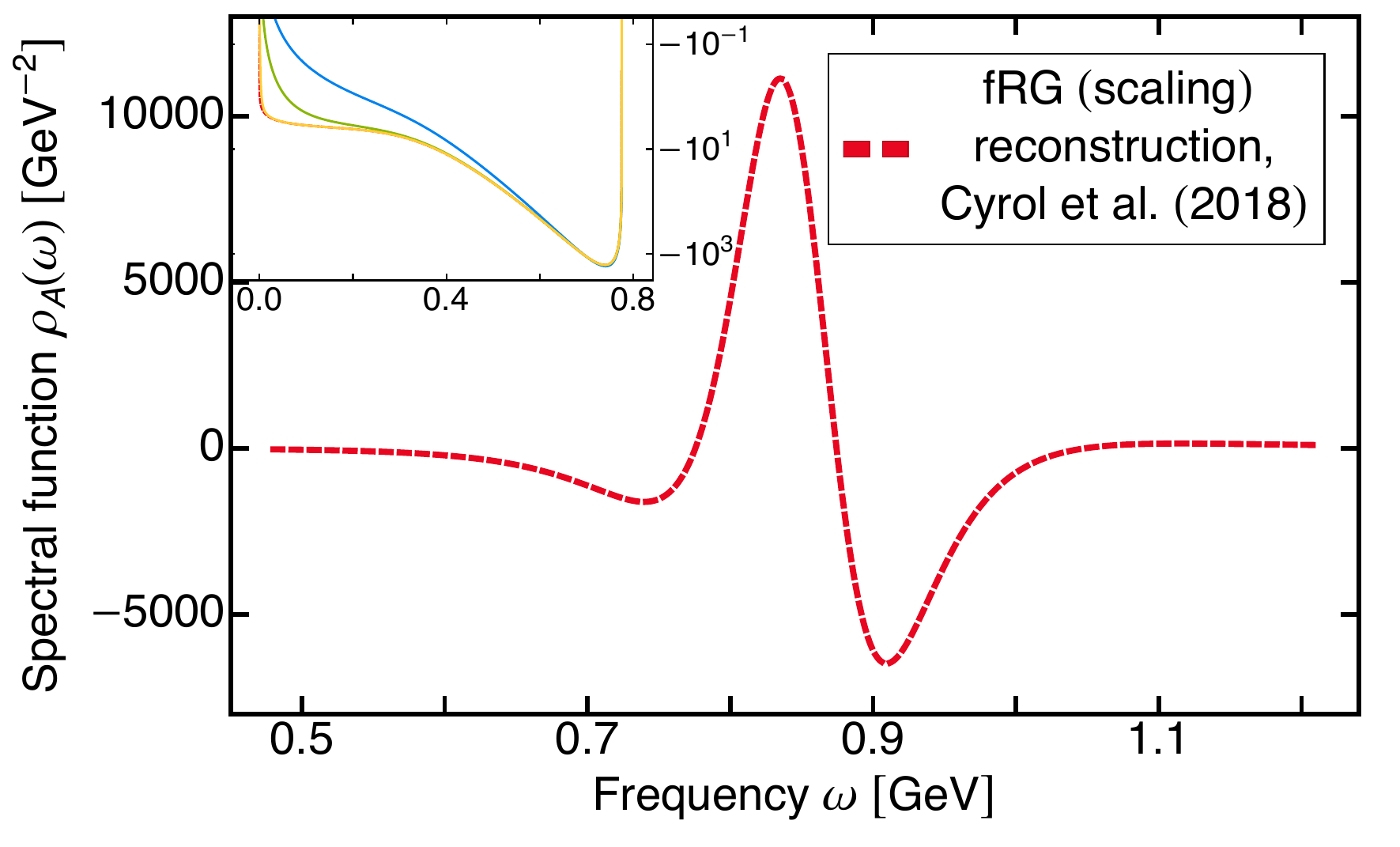}}\hspace{.3cm}
	\subfloat[Euclidean gluon propagators obtained from the KL-representation \labelcref{eq:KL} with the gluon spectral functions in 
	\Cref{fig:gluSpecPlot}. The small IR-difference for $\omega\lesssim 0.7$\,GeV shown in the inset in \Cref{fig:gluSpecPlot} translate into the IR-differences for $p\lesssim 1$\,GeV. The lattice data is taken from~\cite{Sternbeck:2006cg}. \hspace*{\fill}]{
		\includegraphics[width=.48\textwidth]{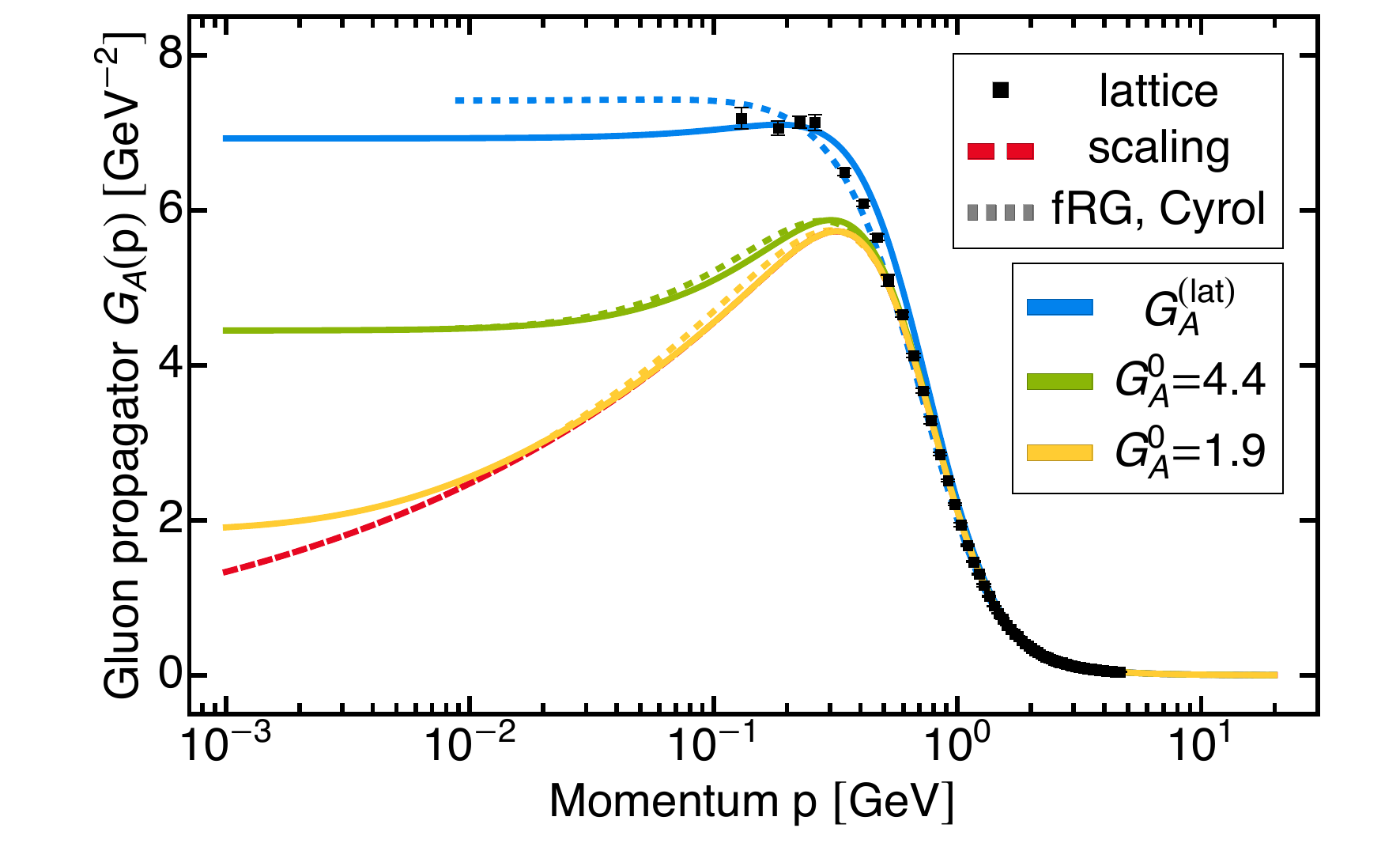}\label{fig:gluPropDecPlot}}
	\caption{Reconstructed gluon spectral function (left) based on~\cite{Cyrol:2018xeq} and \labelcref{eq:gluon_spec_dec} and the respective gluon propagators (right). The spectral functions and propagators differ in the infrared and are labelled by $G_A^\text{(lat)}$ (blue) for our lattice-type input, and $G_A(0)$\,[GeV${}^{-2}$]=4.4 (green), 1.9 (yellow).}
	\label{fig:gluon_input}
\end{figure*}
After the integration over the loop momentum is carried out, we are left with two spectral integrals. They suffer from the (same) logarithmic divergence as the momentum integral, even if simply dropping the $1/\epsilon$-term, that arises from the loop integral. This is a generic feature in the spectral DSE, for a thorough discussion see \cite{Horak:2020eng}. Moreover, the potentially divergent terms in $\Sigma_{\bar c c}(p)$ are proportional to $p^2$. Such a divergence can be cured by a $\tilde Z_3$, that is proportional to $\Sigma_{\bar c c}/p^2$, evaluated at some RG-scale $\mu^{\ }_\text{\tiny{RG}}$. We use a standard renormalisation condition for the (inverse) dressing function,
\begin{subequations}\label{eq:RGcond}
\begin{align}
	\label{eq:renorm_cond}
	Z_c(\mu^{\ }_{\s{\textrm{RG}}}) = \tilde Z_c \,. 
\end{align}
This renormalisation condition is implemented by the respective choice of the renormalisation constant $\tilde{Z}_3$ as 
\begin{align}
	\label{eq:Z3choice}
	\tilde{Z}_3 = \tilde Z_c + \frac{\Sigma_{\bar c c}(\mu^{\ }_{\s{\textrm{RG}}})}{\mu^2_{\s{\textrm{RG}}}}  \,. 
\end{align}
\end{subequations}
In accordance with \labelcref{eq:ghgl-approx}, \labelcref{eq:RGcond} is augmented with $\tilde Z_1\to 1$. For a detailed discussion of self-consistent MOM-type RG-conditions for DSEs, see~\cite{Gao:2021wun}. Eventually, this leads us to the renormalised DSE for the ghost dressing,  
\begin{align} 
	\label{eq:DSEren}
	Z_c(p) = \tilde Z_c - \Bigg[ \frac{\Sigma_{\bar c c}(p)}{p^2}  - \frac{\Sigma_{\bar c c}(\mu^{\ }_{\s{\textrm{RG}}})}{\mu^2_{\s{\textrm{RG}}}} \Bigg] \,.
\end{align}
The explicit choice of the renormalisation condition $\tilde Z_c$ will be discussed in~\Cref{sec:results}.

\subsection{Iterative solution} \label{sec:SolIt}

The renormalised DSE in \labelcref{eq:DSEren} can be evaluated \textit{analytically} for general complex frequencies. For the extraction of the spectral function with \labelcref{eq:spec_func_def} we choose $p_0=-\text{i} (\omega+\text{i} \varepsilon)$ with $\varepsilon \to 0$. This leads us to
\begin{align} 
	\label{eq:DSE_realtime}
	Z_c(\omega) = \tilde Z_c + \Bigg[ \frac{\Sigma_{\bar c c}(\omega)}{\omega^2} + \frac{\Sigma_{\bar c c}(\mu^{\ }_{\s{\textrm{RG}}})}{\mu^2_{\s{\textrm{RG}}}} \Bigg] \,,
\end{align}
where, in a slight abuse of notation, we define ${\Sigma_{\bar c c}(\omega) = \Sigma_{\bar c c}(-\imag (\omega + \imag 0^+))}$. Note that the $1/(\omega+\imag \epsilon)^2$ term does not trigger a $\delta$-function, as the self energy is proportional to $(\omega+\imag \epsilon)^2$ for small $\omega$. This allows us to use $1/(\omega+\imag \epsilon)^2\to 1/\omega^2$, as is done in \labelcref{eq:DSE_realtime}. We also check at every iteration step that no further poles are generated, and hence \labelcref{eq:DSE_realtime} holds true.

The explicit spectral integral expression for $\Sigma_{\bar c c}(\omega)$ and its renormalised counter part can be found in~\Cref{app:loop_mom_int}. The remaining finite spectral integrals have to be computed numerically, and the spectral function $\rho_c(\omega)$ is given with \labelcref{eq:spec_func_def} as 
\begin{align}\label{eq:rhocG}
	\rho_c(\omega)= \frac{\pi}{Z_c}\delta(\omega^2) - \frac{2}{\omega^2} \, \textrm{Im} \Big[ \frac{1}{\, Z_c(\omega)} \Big] \,. 
\end{align}

The substitution of \labelcref{eq:DSE_realtime} into \labelcref{eq:rhocG} allows us to compute the ghost spectral function as well as the propagator for Euclidean and time-like frequencies, for a given gluon spectral function $\rho_A$. This input is discussed in \Cref{sec:GluonSpec}. The ghost spectral function is then computed within the following iteration procedure, discussed in detail in~\cite{Horak:2020eng}, and briefly described here. 

The ghost spectral function $\rho_c^{(i)}$, obtained after the $i$-th iteration step, and the given input $\rho_A$, are inserted into the spectral integral form of $\Sigma_{\bar c c}(p)$, on the right-hand side of \labelcref{eq:DSE_realtime}. Then, by means of \labelcref{eq:rhocG}, we arrive at the ($i+1$)-th spectral function, $\rho_c^{(i+1)}$. This iteration is repeated until convergence has been reached. 

The iteration commences with an initial guess for $\rho_c$. Here we choose the canonical choice of its ``classical'' spectral function, \ie, a massless pole with residue one,
\begin{align}
	\label{eq:initial_guess_ghost}
	\rho_c^{(0)}(\omega) = \pi \, \delta(\omega^2) \,.
\end{align}
This particular choice leads to a stable and rapid convergence of the iteration procedure. For a discussion of the numerics and their convergence properties, see \Cref{app:numerics} and \Cref{fig:convergence_plot}.

\subsection{Gluon spectral function} \label{sec:GluonSpec}

As already mentioned, the input gluon spectral function is taken from~\cite{Cyrol:2018xeq}, where a spectral reconstruction of the Yang-Mills gluon propagator fRG data from~\cite{Cyrol:2016tym} has been performed. In both scaling and decoupling scenarios, the infrared behaviour of the gluon spectral function (assuming the validity of the KL representation) can be inferred from the respective infrared scaling of the gluon propagator in \labelcref{eq:gluon_prop_param}. More details can be found in~\cite{Cyrol:2018xeq}. 

For the entire family of solutions, the deep infrared limit with $p\to 0$ is parametrised by~\cite{Cyrol:2018xeq}, 
\begin{subequations} \label{eq:IR-Family} 
\begin{align} 
	\label{eq:GA-Family} 
	\hat G_A(p) =  & \, Z_A^{(\textrm{IR})} \, x^{-1+2 \kappa}  \,,
\end{align}
with a constant $Z_A^{(\textrm{IR})}$. The scaling coefficient $\kappa$ takes values in the range $1/2 < \kappa < 1$, and 
\begin{align}\label{eq:x}
	x= {\hat p}^{\,2}+\gamma_{G} \left(\hat m_\textrm{gap}^2+ {\hat p}^{\,2}\log {\hat p}^{\,2}\right)\,,
\end{align}
\end{subequations}
where the hatted dimensionless quantities in \labelcref{eq:GA-Family} all future expressions have been rescaled with the appropriate powers of $\Lambda_\textrm{QCD}$, \eg, ${\hat p}^{\,2} = p^2/\Lambda_\textrm{QCD}^2$. For \mbox{$\gamma_{G}=0$}, the gluon propagator in \labelcref{eq:GA-Family} reduces to the scaling propagator. 

The lattice-type propagator is obtained for a $\gamma^{(\textrm{lat})}_G$ that is close to the maximal one compatible with infrared QCD in the Landau gauge. 
The parameters $(\gamma_G, \hat m^2_{\textrm{gap}})$ characterise the one-parameter family of solutions. Indeed, the actual solutions in~\cite{Cyrol:2016tym} are well approximated by using the functional form of the scaling solution but 
with the argument of \labelcref{eq:x}, and an appropriate tuning of $\gamma_G$. 
We shall exploit this property for constructing a simple one-parameter family of gluon spectral functions,
using the scaling one, $\rho^{(\textrm{dec})}(\omega)$, as our point of departure.

For completeness,  we note that, for $p\to 0$, the respective (dimensionful) ghost propagator is given by 
\begin{align}
	G_c(p)  = & \, \frac{Z_{c}^{(\textrm{IR})}}{p^2}   \frac{1}{x^{2 \kappa}} \,,
\end{align}

In the deep infrared, $\rho^{(\textrm{scal})}(\omega)$ is determined from \labelcref{eq:GA-Family} and \labelcref{eq:x}, setting $\gamma_G=0$. Specifically, for $\omega\to 0_+$, we obtain 
\begin{align}\label{eq:rhoAIR-scale}
 \hat\rho^{\,(\textrm{scal})}_A(\omega) = &  -2\,Z_A^{(\textrm{IR})}\,{\hat\omega}^{2 (2\kappa-1)}\,, 
\end{align}
which corresponds to the infrared tail of the full spectral function reconstructed~\cite{Cyrol:2018xeq}, depicted in \Cref{fig:gluon_input}. Similarly, in the case of the decoupling-type solutions, we arrive at  
\begin{figure*}[t]
\centering
		\subfloat[Ghost spectral function: direct computation by iteration with the spectral DSE \labelcref{eq:ghost_self_energy_final} and the gluon spectral function from \Cref{fig:gluon_input} for $G_A^\text{(lat)}$. The inlay also indictaes the $\delta$-function contribution in the origin, indicated by an arrow. Its  amplitude is given by the value of corresponding Euclidean dressing function $1/Z_c(p)$ at $p=0$. The squares show our best fit, comp.~\Cref{sec:spectral_fits}. \hspace*{\fill}]{\includegraphics[width=.48\textwidth]{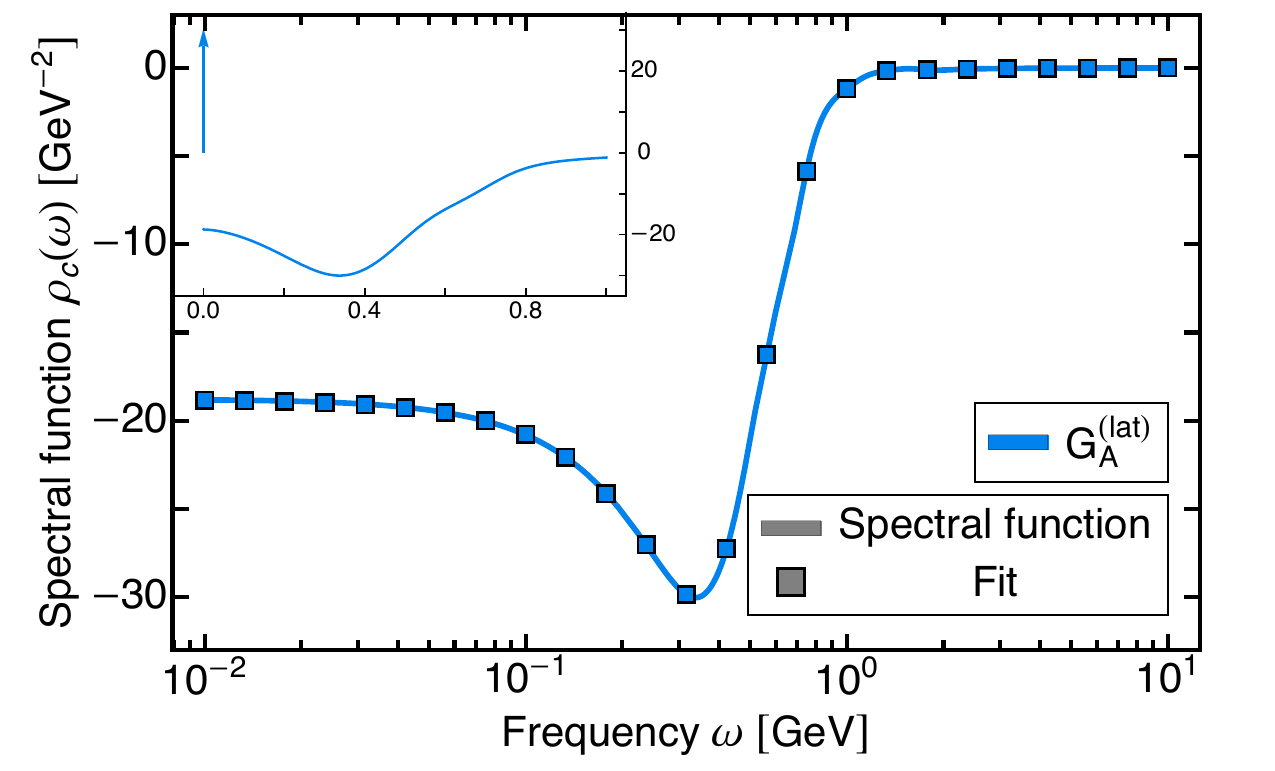}\label{fig:ghostSpecPlot}} \hspace{.2cm}
		\subfloat[Euclidean ghost dressings: $1/Z_c(p)$ from KL-representation via the $\rho_c$'s in \Cref{fig:ghostSpecPlot} (squares), $1/Z_c(p)$ from the direct solution with the Euclidean DSE (straight lines), $1/Z_c(p)$ from the Euclidean fRG computations in~\cite{Cyrol:2016tym}: we have taken the solutions with matching values of $G_A^0$ (dashed lines). \hspace*{\fill}]{
		\includegraphics[width=.48\textwidth]{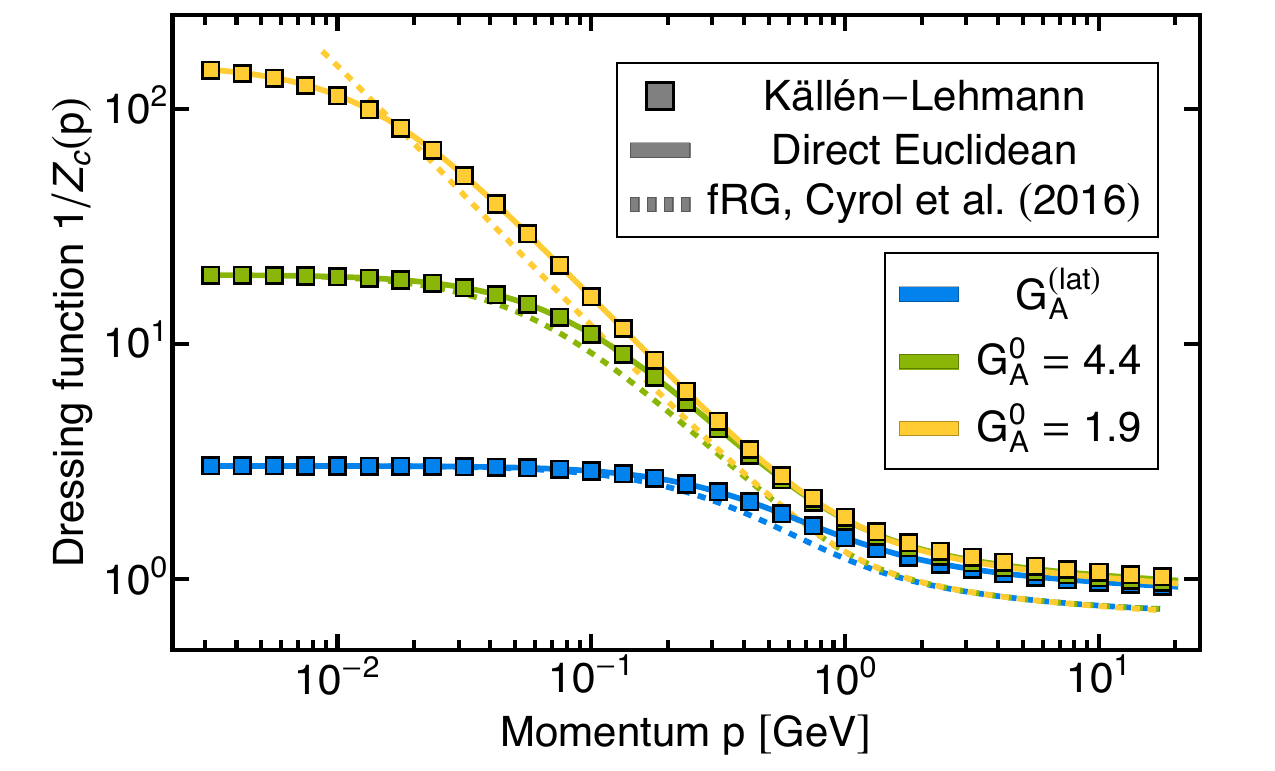}\label{fig:ghostDressPlot}}
	\caption{Ghost spectral functions (left) and respective Euclidean dressings (right), obtained with the decoupling gluon spectral functions in \Cref{fig:gluon_input}, with the same colour coding by $G_A^\text{(lat)}$ (blue), $G_A(0)$\,[GeV${}^{-2}$]=4.4 (green), 1.9 (yellow).}
	\label{fig:ghost_spec_dress}
\end{figure*}
\begin{align}\label{eq:rhoAIR-dec} 
 \hat \rho_A^{(\textrm{dec})}(\omega) = &  -\,\frac{Z_A^{(\textrm{IR})}}{\gamma_G} \,\frac{2 \pi }{\hat m^4_\textrm{gap}} \hat \omega^2\,.
\end{align}
While \labelcref{eq:rhoAIR-scale} and \labelcref{eq:rhoAIR-dec} describe the different behaviour of the scaling and decoupling spectral functions in the deep infrared, for larger spectral values the two sets of spectral functions coincide. This regime is approximately bounded from below by the first zero, $\omega_0$, of the scaling spectral function, shown in \Cref{fig:gluon_input}, with $\omega_0\approx 0.78$. A simple interpolation to the decoupling solution, based on the scaling spectral function in~\cite{Cyrol:2018xeq}, is therefore given by 
\begin{subequations}
	\label{eq:gluon_spec_dec}
\begin{align}\nonumber 
	\rho_A^{(\text{dec})}(\omega,\chi) =&\, Z_\chi \left( \frac{\omega^2} {\omega^2 + \chi^2}\right)^{2-2\kappa }\rho_A^{(\text{scal})}(\omega)\theta(\omega_0-\omega)\\[2ex] & \hspace{1cm}
	+ \rho_A^{(\text{scal})}(\omega)\theta(\omega-\omega_0)\,,
\label{eq:gluon_spec_mod} 
\end{align}
with 
\begin{align}\label{eq:Zchi} 
	Z_\chi := \frac{  \int_0^{\omega_0} d\lambda\, \lambda\, \rho_A^{(\text{scal})}(\lambda)}{\int_0^{\omega_0} d\lambda \,\lambda\, \rho_A^{(\text{scal})}(\lambda)
		\Big(  \frac{\omega^2}{\omega^2 + \chi^2} \Big)^{2-2\kappa} }\,,
\end{align}
\end{subequations}
dictated by the Oehme-Zimmermann superconvergence relation for the gluon spectral function, 
\begin{align}\label{eq:OZ-gluon}  
 \int_0^\infty d\lambda \, \lambda \, \rho_A(\lambda)=0\,. 
 \end{align}
With \labelcref{eq:Zchi} the total spectral weight of $\rho_A^{(\text{dec})}(\omega,\chi)$ is the same as that of $\rho_A^{(\text{scal})}(\omega)$. Hence, given that the latter satisfies \labelcref{eq:OZ-gluon}, so does the former. The scaling spectral function reconstructed in~\cite{Cyrol:2018xeq}, satisfies \labelcref{eq:OZ-gluon} analytically for $\epsilon \to 0^+$ in  \labelcref{eq:spec_func_def}. In the present work we take a small $\epsilon\approx 10^{-7}$ leading to 
\begin{align}
\label{eq:ApproxOEZ}
\frac{	\int d\lambda\,\lambda\,\rho_A(\lambda) }{	\int  d\lambda\,\lambda\,|\rho_A(\lambda)|} \approx 10^{-4}\,. 
\end{align} 
for all spectral functions. For $\chi=0$, we get back the scaling solution with $\gamma_G=0$. The lattice gluon is achieved via~\labelcref{eq:gluon_spec_mod} for $\chi^{\textrm{(lat)}} = \frac{3}{4} \, \mathrm{GeV^2}$ with $Z_\chi^\text{(lat)} = 1.86$. 

We emphasise that fully quantitative gluon spectral functions $\rho_A^{\textrm{(dec)}}$ may be achieved by means of reconstructions. While possible, this is beyond the scope of the present work. Note also that the simple analytic spectral functions $\rho_A(\omega,\chi)$ give semi-quantitative results for the gluon propagators, while at the same time allowing for an analytic access to the relative changes.

\begin{figure*}[t]
	\centering
		\includegraphics[width=.47\textwidth]{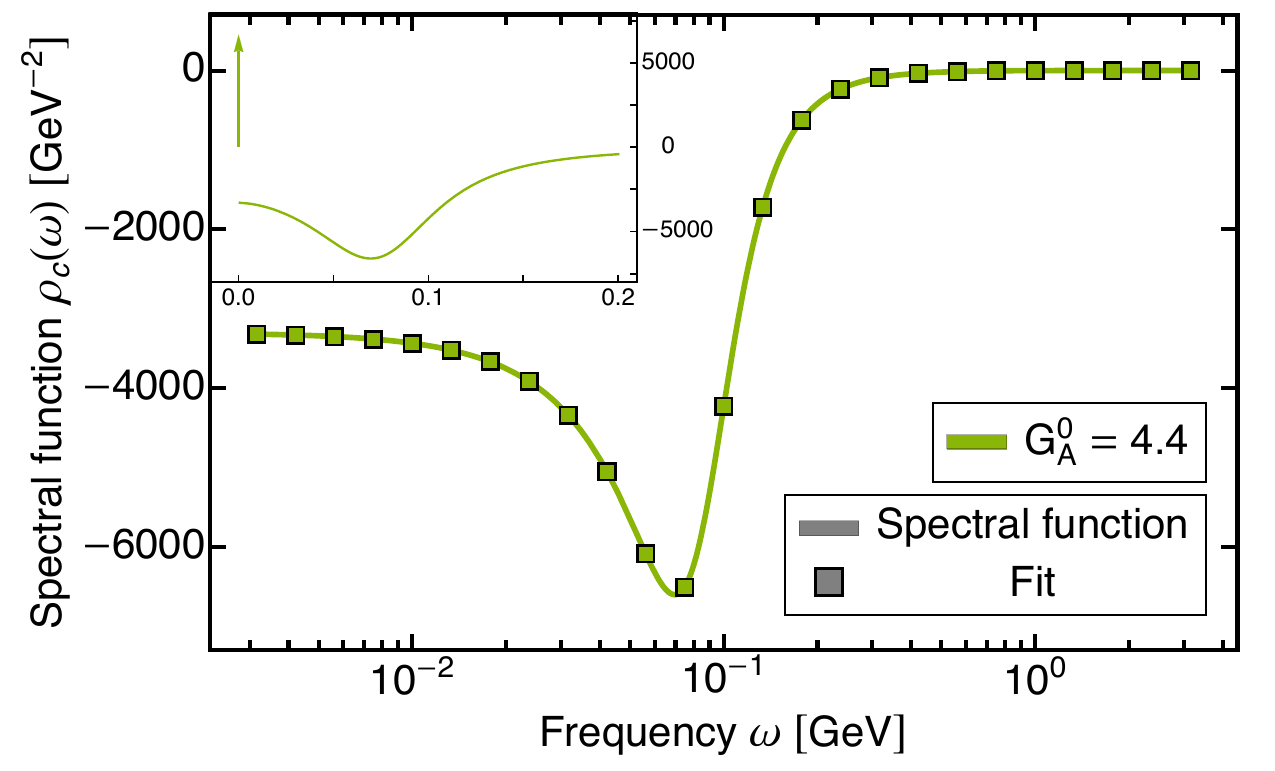}\hspace{.5cm}
		\includegraphics[width=.47\textwidth]{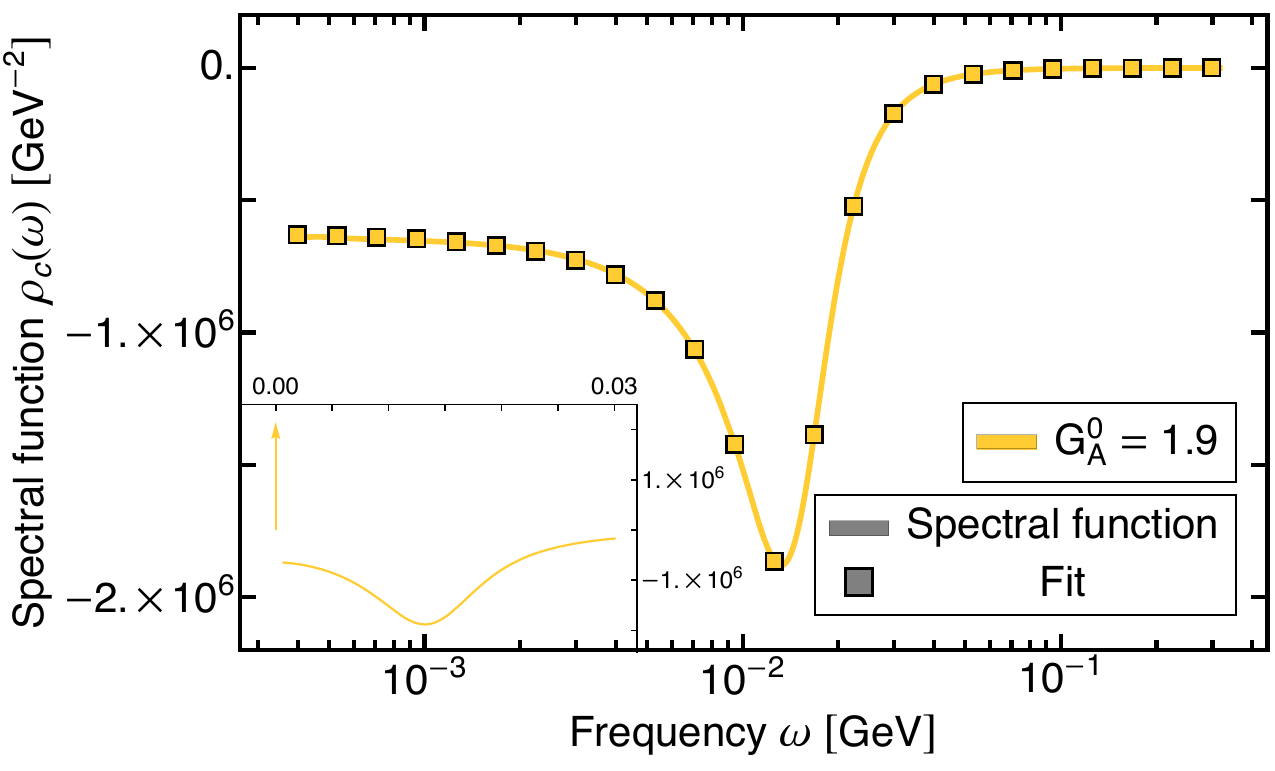}
		\caption{Ghost spectral functions: direct computation by iteration with the spectral DSE \labelcref{eq:ghost_self_energy_final} and the gluon spectral functions from \Cref{fig:gluon_input} with the same colour coding, i.e. $G_A(0) \, [\mathrm{GeV^{-2}}] = 4.4$ (green), 1.9 (yellow). The inlays also show the $\delta$-function pole in the origin, indicated by an arrow. The residue is given by the value of corresponding Euclidean dressing function $1/Z_c(p)$ at $p=0$. The squares show our best fits, comp.~\Cref{sec:spectral_fits}. }
		\label{fig:ghost_specs_23}
\end{figure*}
%


\section{Results}
\label{sec:results}

With the preparation of the previous sections we now compute the ghost spectral function. The ghost DSE is solved for the three different input decoupling gluon spectral functions in \Cref{fig:gluon_input} and propagators, labelled by $G_A^\text{(lat)}$ resp. the infrared value of the related gluon propagators $G_A(0)=4.4, 1.9\,$[GeV${}^{-2}$]. For $G_A^\text{(lat)}$, we tune the mass parameter $\chi$ in~\labelcref{eq:gluon_spec_mod} such that we best agree with the lattice results from~\cite{Sternbeck:2006cg}. We pair each of our input $G_A$'s with a gluon propagator from the family of self-consistent YM solutions of~\cite{Cyrol:2016tym}, indicated by dashed lines in the right panel of~\Cref{fig:gluon_input}. For $G_A^\text{(lat)}$ (blue curve), we chose the solution which also matches the lattice results from~\cite{Sternbeck:2006cg} best. The green and yellow curves are matched with the respective solution with the same $G_A(0)$. 

The renormalisation condition $\tilde Z_c$ is now chosen such that the value of the ghost dressing function $1/Z_c(p)$ matches that of~\cite{Cyrol:2016tym} at the RG scale $\mu^{\ }_\text{\tiny{RG}} = 20$ MeV for the respective gluon propagator. This IR renormalisation procedure is necessary in order to compensate for the lack of self-consistency when considering the ghost DSE with fixed gluon input. The strong coupling constant is fixed to $\alpha_s = 0.26$.

In \Cref{fig:ghostSpecPlot} and \Cref{fig:ghost_specs_23} we show the respective ghost spectral functions. All spectral functions shows a positive particle peak at vanishing momentum, constituted by a delta distribution. The magnitude of the corresponding residue, i.e. the particle peaks amplitude, rises with decreasing $G_A(0)$, and the residues positivity reflects the chosen positive classical dispersion of the ghost. The spectral function also has a negative scattering spectrum starting at vanishing frequency. For decreasing $G_A(0)$, one gradually approaches the scaling solution, and the spectral weight increases drastically. This also mirrors the increasing amplitude of the particle peak, which is enforced by the Oehme-Zimmermann-type superconvergence relation \labelcref{eq:ghost_spec_norm} for the ghost, for more details see \Cref{app:sum_rules} and \cite{Bonanno:2021squ}. This also leads to the known UV-asymptotics for the ghost spectral function, 
\begin{align} \label{eq:rho_uv}
	\hat\rho^\text{(UV)}(\hat\lambda) = \frac{Z_c^{\textrm{(UV)}}}{1+\hat\lambda^2 (\log \hat\lambda^2)^{\gamma_c}} \,,
\end{align}
with the ghost anomalous dimension $\gamma_c$, and the UV wave function renormalisation $Z_c^{\textrm{(UV)}}$. The gluon spectral function for $G_A^\text{(lat)}$ represents the lattice-type case, see \Cref{fig:gluPropDecPlot}. The respective lattice data for the ghost propagator is depicted in \Cref{fig:ghostDressPlot}, and confirms the semi-quantitative nature of the classical vertex approximation in the ghost DSE. For smaller $G_A(0)\to 0$, the gluon propagator approaches the scaling solution. This entails, that also the ghost propagators approaches the scaling solution with $1/Z_c(p)\propto (p^2)^{-\kappa}$.

The Euclidean dressing functions corresponding to the computed spectral functions for the different gluon propagator inputs are shown in~\Cref{fig:ghostDressPlot}. We show both the dressing functions obtained from the spectral Euclidean and real-time DSE and find that the spectral representation for the ghost propagator (and dressing function) holds. We also compare to the Yang-Mills results from~\cite{Cyrol:2016tym}, which we also used in the renormalisation condition as described above, and see that we reach very good qualitative agreement. In particular, our most scaling-like solution shows the typical scaling behaviour down to about 30 MeV. The deviations from~\cite{Cyrol:2016tym} in the UV originate in the  classical approximation for the full ghost-gluon vertex used here.

\begin{figure}[b]
	\centering
	\includegraphics[width=.46\textwidth]{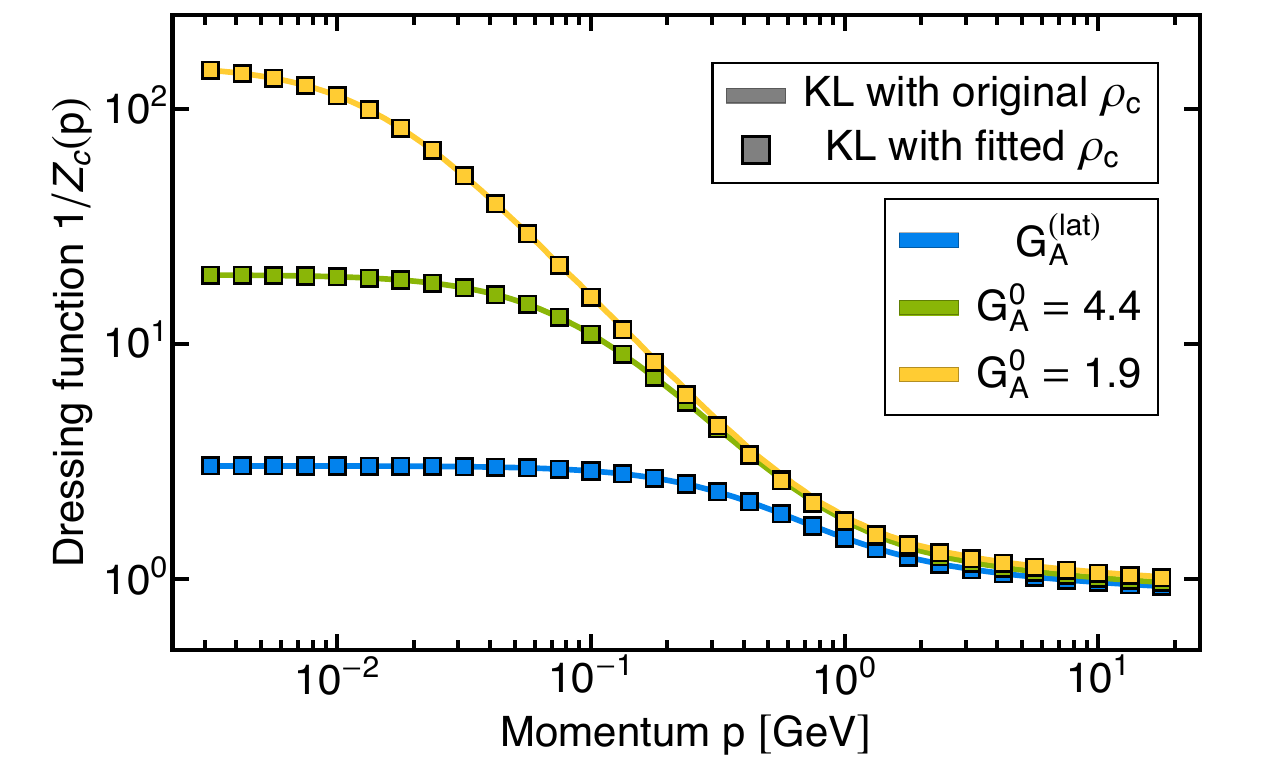}
	\caption{Comparison of the ghost dressing function obtained via the KL representation~\labelcref{eq:tilderho} from the spectral function (solid line) and its fit (squares) for the different input gluon propagators.}
	\label{fig:fit_plot}
\end{figure}
%
\subsection{Comparison with previous works} \label{subsec:comparison}
\begin{figure*}[t]
	\centering
	\includegraphics[width=0.46\linewidth]{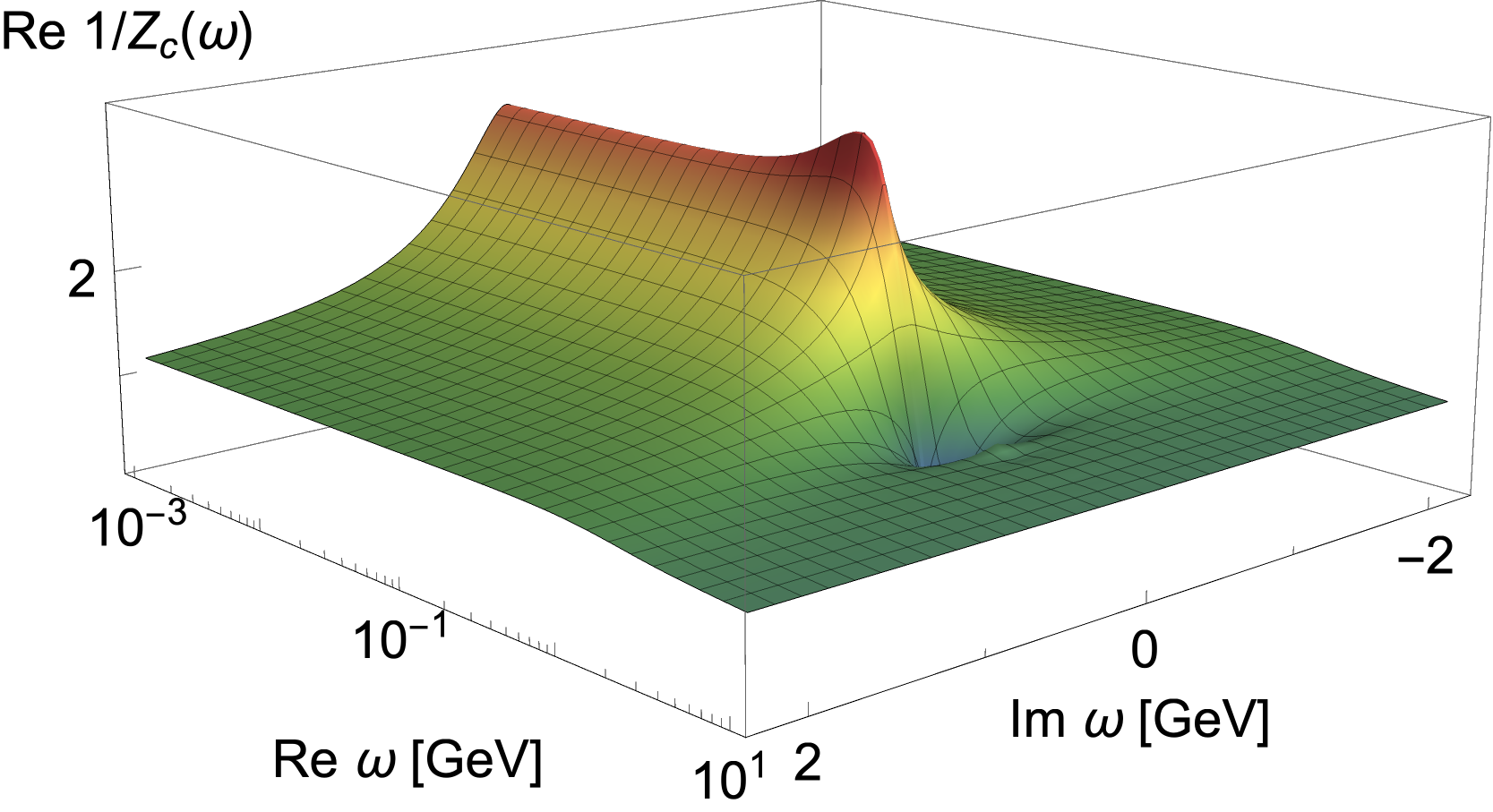}\hspace{1cm}
	\includegraphics[width=0.46\linewidth]{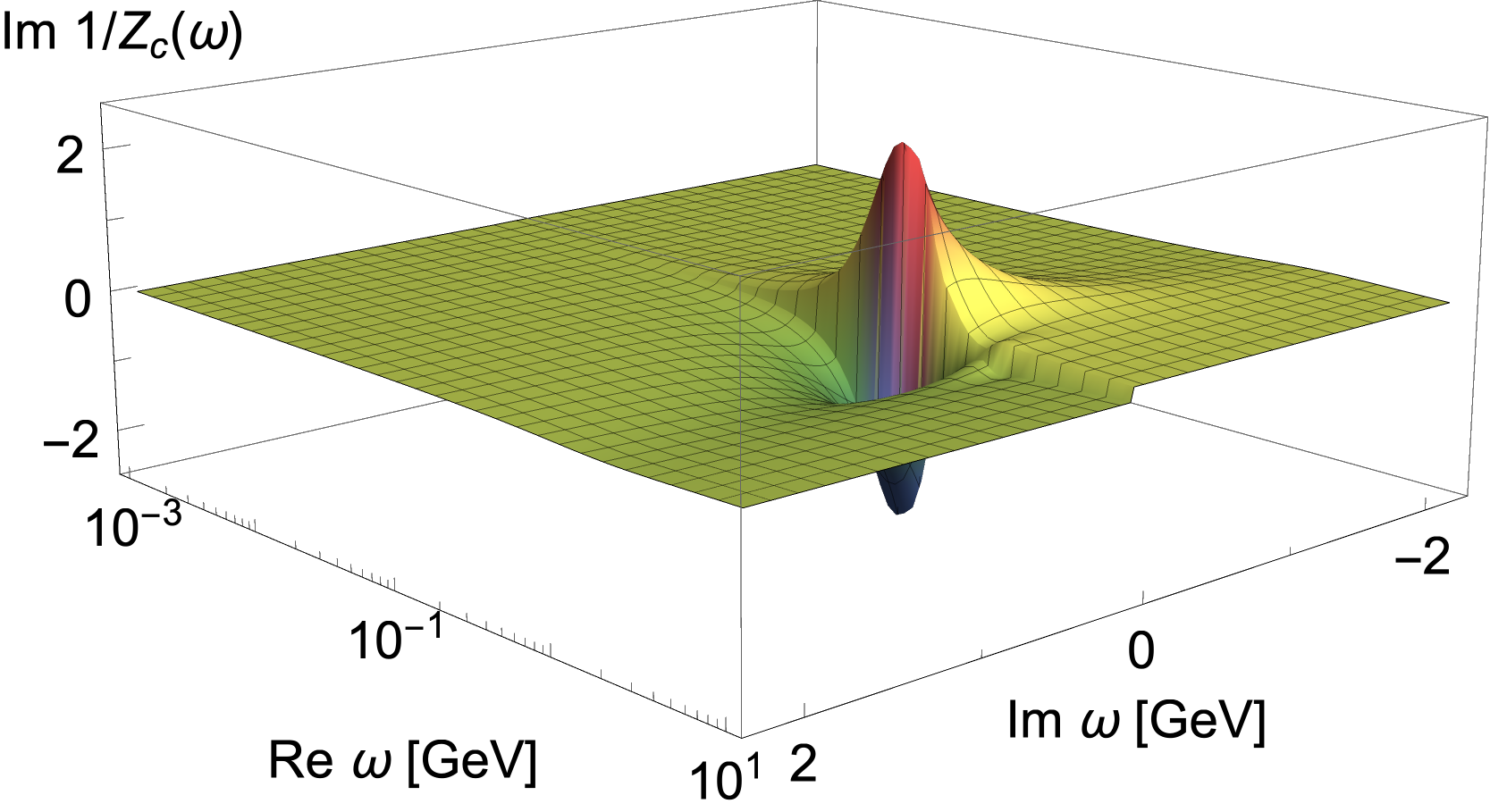}
	\caption{Real (left) and imaginary (right) part of the ghost dressing function $1/Z_c(\omega)$ for the lattice-like gluon input (comp.~\Cref{fig:ghost_spec_dress}) as a function of complex frequencies. Purely real $\omega$ correspond to Minkowski frequencies, purely imaginary $\omega$ to Euclidean frequencies. The colour coding serves to guide the eye. The branch cut along the real frequency axis is clearly visible.}
	\label{fig:ghost_spec_dress_comp_plane}
\end{figure*}
\begin{figure}[b]
	\centering
	\includegraphics[width=\linewidth]{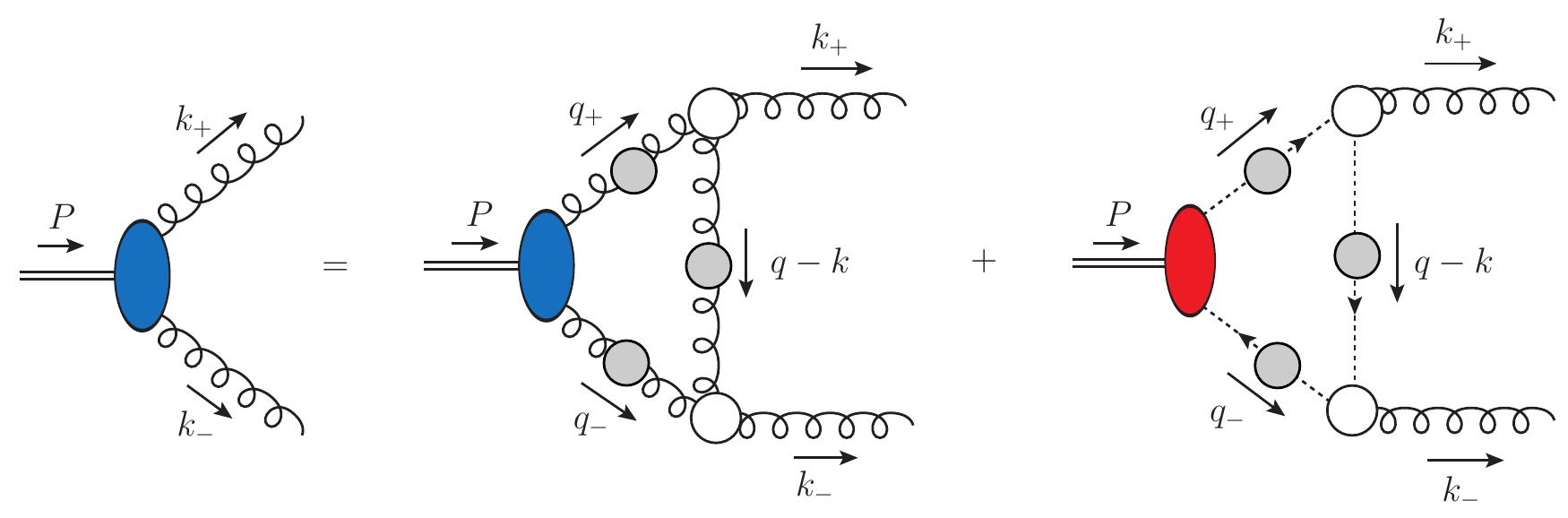}
	\caption{One of the two BSEs comprising the system that controls the scalar glueball formation.
		The blue (red) ellipses denote the glueball-gluon (ghost) BS amplitudes, and   
		\mbox{$k_{\pm} = k \pm P/2$}, \, $q_{\pm} = q \pm P/2$, where $q$ denotes the loop momentum. }
	\label{fig:BSE}
\end{figure}

In this section we compare our results on the ghost spectral function with that in the literature, for results with  different approaches see \cite{Siringo:2016jrc, Binosi:2019ecz, Dudal:2019gvn, Falcao:2020vyr}. The spectral function also allows us to map out the ghost propagator in the complex momentum plain, which is discussed in \Cref{subsec:complex_plane} including a comparison with respective results in the literature from a DSE analysis, see \cite{Fischer:2020xnb}. 

In~\cite{Dudal:2019gvn}, the ghost spectral function has been reconstructed from lattice QCD data. The results are in good agreement with our direct computation: both show a massless particle pole and a negative scattering tail. The reconstruction in~\cite{Dudal:2019gvn} lacks reliability for spectral values smaller than roughly $100$\, MeV, as the smallest Euclidean data point used for the reconstruction is at about $p=150$ MeV. In this regime, the present results from a direct spectral computation can be used as an input for future reconstructions by restricting the respective infrared completion. The same qualitative features are also found in the ghost spectral function obtained via a massive propagator expansion in~\cite{Siringo:2016jrc}, i.e. a  massless particle pole as well as negative spectral tail.

In~\cite{Binosi:2019ecz}, Pade-type reconstructions of the ghost dressing spectral function from DSE and lattice data in Yang-Mills theory has been performed. These results are in contradistinction to the present result and~\cite{Dudal:2019gvn, Siringo:2016jrc}, as the scattering tails in~\cite{Binosi:2019ecz} show significant negative contributions. This corresponds to positive contributions in the propagator spectral function due to a relative sign in the definition. In addition, the UV tail of the reconstruction of DSE data is not shown in~\cite{Binosi:2019ecz}, and the spectral function appears to approach a constant value. A UV-positive (negative) as well as a non-vanishing tail in the propagator (dressing) spectral function violates the analytically given asymptotic fixed by~\Cref{eq:spec_func_def}, for a detailed derivation see~\Cref{app:sum_rules}. 

The study of the analytic structure of the ghost propagator put forward in~\cite{Falcao:2020vyr} also suggests the existence of a massless pole as well as a branch cut along the real frequency axis. As already mentioned above, Yang-Mills propagators in the whole complex momentum plane have been investigated with DSEs in~\cite{Fischer:2020xnb}. The findings show good qualitative agreement with the propagators obtained from the ghost spectral function computed in the present work, but do not support a KL spectral representation of the ghost. This is discussed further in~\Cref{subsec:complex_plane}.

\subsection{Spectral fits} \label{sec:spectral_fits}

The results for the ghost spectral function with the UV asymptotics $\rho^\text{(UV)}(\lambda)$ in \labelcref{eq:rho_uv}, the IR asymptotics $\rho_0$ allow for a simple fit in terms of the both asymptotics and Breit-Wigner functions for intermediate spectral values. The split into these three regimes allows for a simple parametrisation $\rho^\text{(fit)}_c$ of the ghost spectral function, 
\begin{subequations}\label{eq:spec_fit}
\begin{align} \nonumber 
	\rho_c^\text{(fit)}(\lambda) & \, = \kappa \Bigg[ \hat\rho_0 \, \sigma_\text{IR}(\lambda) + \, \sigma_1(\lambda) \hat f_\text{peak}^\text{(BW)}(\lambda) \sigma_2(\lambda)\\[1ex]
	& \, + \sum_j^N \hat f_j^\text{(BW)}(\lambda) + \sigma_\text{UV}(\lambda) \hat\rho^{\text{(UV)}}(\lambda) \Big] \,.
	\label{eq:rho_cfit}
\end{align} 
In \labelcref{eq:rho_cfit} we use the sigmoid function for projecting on the three regimes, 
\begin{align} \label{eq:sigmoid}
	\sigma_x(y) = \frac{1}{1+e^{-\nu_x(y-\Lambda_x)}} \,.
\end{align}
where $\kappa$ only carries the appropriate dimension. The intermediate regime is expanded in Breit-Wigner kernels, 
\begin{align} \label{eq:breit_wigner}
	\hat f_x^\text{(BW)}(\hat y) = \frac{c_x}{(\hat y- \hat M_x)^2 + \hat\Gamma_x} \,.
\end{align}
\end{subequations}
For our best fit, we use $N=3$. The respective fit parameters are listed in~\Cref{app:numerics},~\Cref{tab:fit_params}, and the fits are depicted together with the spectral functions in \Cref{fig:specs_log_log}. 

The accuracy of the fits is best evaluated within a comparison between the ghost dressing functions $1/Z_c(p)$ obtained from the computed spectral functions and their fits. This comparison if depicted in~\Cref{fig:fit_plot} for all three different input gluon propagators. 

\subsection{Results in the complex plane} \label{subsec:complex_plane}

We close this section with a short discussion of the potential application of the present results within bound state and resonance computations in QCD. To begin with, the behaviour of QCD correlation functions for complex-valued momenta is instrumental for the reliable computation of bound-state properties within the frameworks of the Bethe-Salpeter equations (BSEs). In this quest, the gluon and ghost propagators are of paramount importance, as may be exemplified by considering the BSEs that control the formation of glueballs in a pure Yang-Mills theory~\cite{Dudal:2010cd, Meyers:2012ka, Meyer:2015eta, Sanchis-Alepuz:2015hma, Souza:2019ylx, Huber:2020ngt} (for lattice studies, see~\cite{Athenodorou:2020ani} and references therein). In fact, the present results are specifically useful for the scalar glueball: in contradistinction to its pseudo-scalar counterpart, it involves {\it both} the gluon and ghost propagators, as shown in~\Cref{fig:BSE}.

As is well-known, the need to extend the aforementioned propagators to the complex plane stems from the fact that the momentum $P$ of the bound-state in question must satisfy $P^2=-{M}^2$, where ${M}$ is the corresponding mass. This condition is typically implemented by introducing the rest-frame parametrization $P = {M} (0,0,0,\mathrm{i})$ (see, \eg,~\cite{Sanchis-Alepuz:2015tha}). Invariably, this complexifies the arguments of $G_A(q_{\pm})$ and $G_c(q_{\pm})$ in the BSE of~\Cref{fig:BSE},  since $q^2_{\pm} = \left|q \right|^2 - {M}^2/4  \pm  \mathrm{i}\left|q\right| {M} $. 

These considerations motivate the computation of the dressing function $1/Z_c(p)$ in the entire complex plane. To that end, we employ the KL representation of \labelcref{eq:spec_rep_dressing}, utilising the $\rho_c(\lambda)$ found above, and setting \mbox{$ \imag p= {\mathrm Re}\, \omega + \mathrm{i}\, {\mathrm Im} \,\omega$}. The results of this computation are shown in \Cref{fig:ghost_spec_dress_comp_plane}. 

We now compare our results for the ghost propagator  in the complex plane with the spectral DSE with those from~\cite{Fischer:2020xnb}. There, ghost and gluon propagators in the complex plane have been computed with complex DSEs. The gluon propagators in~\cite{Fischer:2020xnb} exhibit complex conjugate singularities, and their nature and position varies greatly under small changes in the ansatz for the vertex. We emphasise, that these singularities simply indicate the limited radius of convergence of the method both for the gluon and for the ghost, for a detailed discussion see~\cite{Fischer:2020xnb}. For large (angular) distances to the Euclidean axis analyticity is lost, and the method used does not produce reliable results. If reconstructed with the Schlessinger point method, the singularities observed in~\cite{Fischer:2020xnb} take the form of complex conjugate poles. This has also been seen in \cite{Binosi:2019ecz}, where similar reconstruction methods have been used. For further studies of the complex structure of QCD-like theories in the presence of complex conjugate poles see also the recent work~\cite{Kondo:2019ywt,Hayashi:2020few}. 

Despite the lack of reliability for sufficiently large Minkowski frequencies, we have compared ghost dressing from~\cite{Fischer:2020xnb} with the present result in this region. The imaginary part of the ghost dressing function computed there is strictly positive for time-like momenta, in qualitative agreement with our result, in particular in view of the different approximations. We have also confirmed the absence of a spectral representation of the ghost by computing the spectral function from the Minkowski dressing of~\cite{Fischer:2020xnb} via~\Cref{eq:spec_func_def}. Then, the Euclidean dressing is computed via the KL representation of~\Cref{eq:KL} and compared to the direct calculation. This comparison showed a significant violation of the spectral representation especially for larger Euclidean frequencies. This is to be expected, since by nature of the kernel of the spectral representation~\Cref{eq:KL}, large Euclidean frequencies are sensitive to large spectral values, i.e. large Minkowski frequencies. These lie beyond the radius of convergence of the method used in~\cite{Fischer:2020xnb}, as discussed. 

In summary, this analysis strongly suggests that complex conjugate poles as well as other non-analyticities in the gluon propagator beyond the real frequency axis invalidate the KL representation for both the gluon \textit{and} the ghost. A more detailed discussion is deferred to future work~\cite{horak:preparation}. In particular, this casts serious doubts on mixed reconstructions with a KL representation for the ghost and cc poles for the gluons. In turn, the gluon spectral function in \cite{Cyrol:2018xeq} was reconstructed with the assumption of a KL representation, as outlined in~\Cref{sec:GluonSpec}. As shown in the present work, this also leads to a KL representation of the ghost.  Whether this property holds true in a self-consistent solution of the coupled system, remains to be seen and is deferred to future work. 

\section{Conclusion} \label{sec:Conclusion}

We have solved the Dyson-Schwinger equation for the ghost propagator in the complex plane on the basis of a given input gluon spectral functions, spanning the whole family of decoupling solutions, including the scaling limit. Our spectral DSE approach is based on the spectral DSE put forward in~\cite{Horak:2020eng}, and uses the spectral  renormalisation devised there. The procedure allows for analytic solution of the momentum loop integrals by utilising the KL representation and dimensional regularisation. This facilitates the access to the full complex momentum plane, constituting the central aspect of our scheme. The present truncation uses classical vertices in the ghost gap equation, but we emphasise that the spectral DSE approach also allows for non-trivial vertex approximations, see~\cite{Horak:2020eng}. 

The input data for the gluon spectral function is constructed via a decoupling-type modification of the scaling spectral function from~\cite{Cyrol:2018xeq}. The latter spectral function has been obtained via a reconstruction of the scaling solution fRG data of~\cite{Cyrol:2016tym}. 

The spectral function for the ghost shows a massless pole as well as a continuous scattering tail. The classical massless mode dominates up to momenta close to the position of the maximum of the input gluon propagator. For larger momenta, the perturbative logarithmic behaviour starts to dominate, ultimately causing the dressing function to vanish in the ultraviolet. The present results and the current real-time approach with a real-time renormalisation scheme opens the door to a systematic spectral access to dynamical, time-like properties of QCD. We hope to report on a self-consistent spectral investigation of the full Yang-Mills system soon, both within the standard DSE approach and within the pinch technique. The respective results are pivotal for following studies of the resonance properties and the dynamics of QCD within the present approach. \\[1ex]

\noindent{\bf Acknowledgements}\\[-1ex] 

We thank C.F.~Fischer and M.Q.~Huber for discussions. This work is supported by the Deutsche Forschungsgemeinschaft (DFG, German Research Foundation) under Germany's Excellence Strategy EXC 2181/1 - 390900948 (the Heidelberg STRUCTURES Excellence Cluster) and under the Collaborative Research Centre SFB 1225 (ISOQUANT) and the BMBF grant 05P18VHFCA, by the  Spanish Ministry of Economy and Competitiveness (MINECO) under grant FPA2017-84543-P, and the  grant  Prometeo/2019/087 of the Generalitat Valenciana. JH acknowledges support by the GSI FAIR project.


\vfill

\bookmarksetup{startatroot}
\appendix

\section{Spectral sum rules from perturbative dressing functions} \label{app:sum_rules}

In a general manner, given a KL representation, the normalisation relation for the corresponding spectral function can be inferred from the perturbative behaviour of the propagator. Multiplying~\labelcref{eq:KL} by $p^2$, one has
\begin{align} 
	\frac{1}{Z(p)} = p^2 \int \frac{d\lambda^2}{\pi} \frac{\rho(\lambda)}{p^2 + \lambda^2} 
	=  \int \frac{d\lambda^2}{\pi} \frac{\rho(\lambda)}{1 + \lambda^2/p^2} \,.
\label{eq:spec_rep_dressing}\end{align}
In the UV, the behaviour of the dressing function $Z(p)$ can be inferred from perturbation theory, $\lim_{p \to \infty} 1/Z(p) = Z_\infty^{-1}$. For large $p^2$, we can also expand the integrand, yielding
\begin{align} \nonumber 
	Z_\infty^{-1} = & \lim_{p \to \infty} \sum_{n = 0}^\infty (-1)^n \int \frac{d\lambda^2}{\pi} \rho(\lambda) \Bigg( \frac{\lambda^2}{p^2} \Bigg)^n \\[1ex] 
	= & \,  \int \frac{d\lambda^2}{\pi} \rho(\lambda) + \lim_{p \to \infty} \Delta(p)  \,,
\label{eq:spec_rep_dressing_expanded}\end{align}
defining
\begin{align} \label{eq:delta}
	\Delta(p) = \sum_{n = 1}^\infty (-1)^n (p^2)^{-n} \int \frac{d\lambda^2}{\pi} \rho(\lambda) \, \lambda^{2n} \,.
\end{align}
We want to show $\lim_{p \to \infty} \Delta(p) = 0$ in order to obtain a normalisation condition for $\rho$ via~\labelcref{eq:spec_rep_dressing_expanded} using the known perturbative asymptotics of the corresponding dressing function. In doing that, we first note that for the spectral integral in the left term of the lower line in~\labelcref{eq:spec_rep_dressing_expanded} to converge, the spectral function must obey
\begin{align} \label{eq:drop_off_spec}
	\lim_{\omega \to \infty} \rho(\omega) \, \omega^2 \log \omega^2 \to 0 \,.
\end{align}
If this requirement does not hold, $\rho$ cannot be normalised in the above form.

Based on the assumption of the existence of above representation~\labelcref{eq:spec_rep_dressing}, $\rho$ can be taken to be integrable on $[0,\infty)$. We choose a scale $\Lambda$ such that for frequencies $\lambda > \Lambda$, $\rho$ is given solely by the leading UV behaviour of its corresponding propagator via~\labelcref{eq:spec_func_def}, see also~\cite{Cyrol:2018xeq}. Denoting the known UV asymptotics as $\rho_\text{UV}$, we then distinguish
\begin{align} \label{eq:rho_split}
	\rho(\omega) = 
	\begin{dcases*}
		\rho_\Lambda(\omega) & if $\omega \leq \Lambda$\,, \\[1ex]
		\rho_{\text{UV}}(\omega) & else\,,
	\end{dcases*}
\end{align}
where $\rho_{\text{UV}}$ now obeys~\labelcref{eq:drop_off_spec}. Note that by the nature of the spectral function being a tempered distribution, it can have distributional contributions such as (higher order) poles. These are allowed in our consideration as long as integrability is not violated. The parametrisation~\labelcref{eq:rho_split} is chosen such that these contributions are contained in $\rho_\Lambda$. We now split the spectral integration interval of~\labelcref{eq:delta} along the split of the spectral function and conclude for finite $\Lambda$ that
\begin{align} \label{eq:sigma_n_lambda}
	 \int_0^\Lambda \frac{d\lambda^2}{\pi} \rho_\Lambda(\lambda) \, \lambda^{2n} < \infty \quad \forall \, n \geq 1 \,,
\end{align}
such that for large momenta, the contribution~\labelcref{eq:sigma_n_lambda} to the spectral representation of the dressing function vanishes,
\begin{align} \label{eq:lim_sigma_n_lambda}
	\lim_{p \to \infty} (p^2)^{-n} \int_0^\Lambda \frac{d\lambda^2}{\pi} \rho_\Lambda(\lambda) \, \lambda^{2n} \to 0 \,.
\end{align}
Hence, in the limit of large $p$ we are only left with spectral integral over $\rho_\text{UV}$ contributing to $\Delta(p)$ in~\labelcref{eq:delta}. Taking into account the known asymptotics of $\rho_\text{UV}$ from~\labelcref{eq:drop_off_spec} however, we find that 
\begin{align} \nonumber
&	\lim_{p \to \infty}  (p^2)^{-n} \int_\Lambda^\infty \frac{d\lambda^2}{\pi} \rho_\text{UV}(\lambda) \thinspace \lambda^{2n} \\[1ex] 
&\hspace{2cm}	  < C \lim_{p \to \infty} (p^2)^{-n} \int_\Lambda^\infty \frac{d\lambda^2}{\pi} \, \frac{\lambda^{2n-2}}{\log \lambda^2} \,,
\label{eq:delta_n_uv}\end{align}
where the lower line can already be anticipated to vanish for arbitrary constants $C$. However, this can also be shown rigorously by noting that upon substitution, the last line of~\labelcref{eq:delta_n_uv} can be reexpressed as the exponential integral function $\text{E}_1$,
\begin{align} \label{eq:exp_integral}
	\int_\Lambda^\infty \frac{d\lambda^2}{\pi} \frac{\lambda^{2n-2}}{\log \lambda^2} = - \frac{1}{\pi} \, \text{E}_1(-n \log \lambda^2) \Big|_\Lambda^\infty \,.
\end{align}
The contribution from the lower integral boundary is finite and thus vanishes in~\labelcref{eq:delta_n_uv}. For the upper limit we utilise the asymptotic expansion of the exponential integral and plug this back into~\labelcref{eq:delta_n_uv}, yielding, while dropping the constant prefactor, 
\begin{align} \label{eq:expansion_exp_integral} \nonumber
	\lim_{p \to \infty} & \, (p^2)^{-n} \, \text{E}_1(-n \log p^2) \\[1ex] \nonumber
	& \, = \lim_{p \to \infty} \, (p^2)^{-n} \frac{e^{n \log p^2}}{-n \log p^2} \sum_{m=0}^{\infty} \frac{m!}{(n \log p^2)^m} \\[1ex]
	& \, = \lim_{p \to \infty} \, \sum_{m=0}^{\infty} \frac{-m!}{(n \log p^2)^{m+1}} \,\, \rightarrow \,\, 0 \,.
\end{align}
In conclusion, recalling~\labelcref{eq:delta}, we arrive at
\begin{align} \label{eq:delta_vanishes}
	\lim_{p \to \infty} \Delta(p) = 0 \,,
\end{align}
which, with~\labelcref{eq:spec_rep_dressing_expanded}, eventually yields the desired normalisation for the spectral function,
\begin{align} \label{eq:spec_func_norm_general}
	\int \frac{d\lambda^2}{\pi} \rho(\lambda) = Z_\infty^{-1} \,.
\end{align}
We thus see that, in the fairly general case where $\rho$ can be normalised via the integral in~\labelcref{eq:spec_func_norm_general}, the normalisation is given by the value of the dressing function at infinity.

\section{Loop momentum integration of the ghost-gluon diagram} \label{app:loop_mom_int}

In this appendix we detail the computation of the ghost self energy diagram $\Sigma_{\bar c c}$. Starting at~\labelcref{eq:ghost_self_energy_final}, we define
\begin{align} \label{eq:app_sigma_def}
	\Sigma_{\bar c c}(p) = g^2 \delta^{ab} C_A \int_{\lambda_1,\lambda_2} \lambda_1 \lambda_2 \rho_A(\lambda_2) \rho_c(\lambda_2) \, I(p,\lambda_1,\lambda_2) \,,
\end{align}
with the now dimensionally regularised momentum integral
\begin{align} \label{eq:app_i_def}
	I(p,\lambda_1,\lambda_2) = \int_q \!\Bigg(p^2 - \frac{(p \cdot q)^2)}{q^2} \Bigg) \frac{1}{q^2+\lambda_1^2} \frac{1}{(p+q)^2 + \lambda_2^2} \,.
\end{align}
The measure is now $\int_q = \int d^dq / (2\pi)^d$. 

\subsection{Momentum integration} \label{subsec:appII_momentum_integration}

Next, we employ partial fraction decomposition
\begin{align}
	\frac{1}{q^2} \frac{1}{q^2 + \lambda^2} = \frac{1}{\lambda^2} \Big( \frac{1}{q^2} - \frac{1}{q^2 + \lambda^2} \Big) \,,
\end{align}
and introduce Feynman parameters, i.e. utilise
\begin{align} \label{eq:feyn_param}
	\frac{1}{AB} = \int_0^1 dx \frac{1}{xA+(1-x)B} \,.
\end{align}
Upon a shift in the integration variable $q \to q -xp$ and after some manipulation, we arrive at 
\begin{align} \label{eq:def_I}
	I(p,\lambda_1,\lambda_2) = \int_{q,x} \sum_{i=0}^2 (q^2)^i \Bigg[ \frac{A_i}{(q^2 + \tilde{\Delta}_1)^2 } + \frac{B_i}{(q^2 + \tilde{\Delta}_2)^2} \Bigg] \,.
\end{align}
with 
\begin{align} \nonumber 
	\tilde{\Delta}_1 = & \, (1-x) \lambda_1^2 + x \lambda_2^2 + x(1-x) p^2\,,\\[1ex]
	\tilde{\Delta}_2 = & \, \tilde{\Delta}_1 - x \lambda_2^2 \, .
\label{eq:delta_def}\end{align}
We will not make all intermediate results explicit, such as giving the full expressions for $A_i$ and $B_i$, which are functions of external momentum $p$, the spectral parameter $\lambda_1$ as well as the Feynman parameter $x$. Ultimately, the complete final result will be stated explicitly.

The momentum integrals are now readily solved via the standard integration formulation, 
\begin{align}\nonumber 
	\int & \frac{d^dq}{(2\pi)^d} \frac{q^{2m}}{(q^2+\Delta)^n} \\[1ex] 
	& \, = \frac{1}{(4\pi)^{d/2}} \frac{\Gamma(m+\frac{d}{2}) \Gamma(n - \frac{d}{2}-m)}{\Gamma(\frac{d}{2})\Gamma(n)} \Delta^{m+d/2-n} \,,
 \label{eq:integration_formula}\end{align}
with $m$ a non-negative and $n$ a positive integer. 

\subsection{Feynman parameter integration} \label{subsec:appII_feynman_parameter_integration}

Reordering the expression in powers of the Feynman parameter $x$ and taking the limit $d \to 4 - 2\varepsilon$, we arrive at 
\begin{align} \nonumber 
	I & (p,\lambda_1,\lambda_2) =  \Bigg( \frac{1}{\varepsilon} + \log \frac{4\pi\mu^2}{e^{\gamma_E}} \Bigg) \sum_{i=0}^3 \frac{\alpha_i - \beta_i}{i+1} \\[1ex]
	& \hspace{.5cm} - \int_x \sum_{i=0}^3 x^i \, \big( \alpha_i \log \tilde{\Delta}_1 - \beta_i \log \tilde{\Delta}_2 \big) + {\cal O}(\varepsilon) \, ,
\label{eq:I_dim_limit}\end{align}
with $\gamma_E$ the Euler-Mascheroni constant. The coefficients $\alpha_i$ and $\beta_i$ do not depend on $x$, and will be given down below. We can solve the Feynman parameter integrals analytically and simplify the first sum to obtain the final result,
\begin{align} \nonumber 
	I (p, \lambda_1,\lambda_2) = &\,\Bigg( \frac{1}{\varepsilon} + \log \frac{4\pi\mu^2}{e^{\gamma_E}} \Bigg) \frac{3}{4}p^2 \\[1ex] 
 &\hspace{.6cm}- \sum_{i=0}^3 \big[ \alpha_i \, f_i - \beta_i \, g_i \big]   \, .
\label{eq:I_final}\end{align}
The coefficients $\alpha_i,\beta_i$ are defined as follows:
\begin{align} \nonumber 
	\alpha_0 = & \, \frac{p^2}{2} \,, \\[1ex] \nonumber
	\alpha_1 = & \, - \frac{p^2 (p^2 - 5\lambda_1^2 + \lambda_2^2)}{2\lambda_1^2} \,, \\[1ex] \nonumber
	\alpha_2 = & \, \frac{3 p^2 (3 p^2 - 2 \lambda_1^2 + 2 \lambda_2^2)}{2\lambda_1^2} \,, \\[1ex] 
	\alpha_3 = & \, - \frac{4 p^4}{\lambda_1^2} \,,
\label{eq:def_alpha_i}\end{align}
and 
\begin{align} \nonumber 
	\beta_0 = & \, 0 \,, \\[1ex] \nonumber
	\beta_1 = & \, - \frac{p^2( p^2 + \lambda_2^2)}{2\lambda_1^2} \,, \\[1ex] \nonumber
	\beta_2 = & \, \frac{3 p^2 (3 p^2 + 2 \lambda_2^2)}{2\lambda_1^2} \,, \\[1ex] 
	\beta_3 = & \, - \frac{4p^4}{\lambda_1^2} \,,
\label{eq:def_beta_i}\end{align}
The functions $f_i$ and $g_i$ carry the branch cuts ultimately giving rise to the spectral function and are defined by integrals over the Feynman parameter $x$ via

\begin{align} \label{eq:def_f_g}
	f_i =  \int_0^1 dx \, x^i \log \tilde{\Delta}_1 \,,\qquad 
	g_i =  \int_0^1 dx \, x^i \log \tilde{\Delta}_2 \,,
\end{align}
yielding 

\begin{widetext}
\begin{align} \nonumber 
	f_0 = & \, \frac{\zeta}{2 p^2} D_\text{cut} + 2 \log \lambda_2 + \frac{p^2 - \lambda_1^2 + \lambda_2^2}{p^2} \log\Big(\frac{\lambda_1}{\lambda_2}\Big) - 2 \,, \\[1ex] \nonumber
	f_1 = & \,  \frac{1}{4 p^4 \zeta}  D_\text{cut} \Big[ \big((\lambda_1-\lambda_2)^2+p^2\big) \big((\lambda_1+\lambda_2)^2 + p^2\big) \big(p^2 - \lambda_1^2 + \lambda_2^2\big) \Big] + \log \lambda_2 - \frac{p^2 - \lambda_1^2 + \lambda_2^2}{2 p^2} \\ \nonumber
	& \, + \frac{(\lambda_1^2-\lambda_2^2)^2 + 2 \lambda_2^2 p^2 + p^4}{2 p^4} \log \Big(\frac{\lambda_1}{\lambda_2}\Big) - \frac{1}{2} \,, \\[1ex] \nonumber
	f_2 = & \, \frac{1}{6 p^6 \zeta} D_\text{cut} \Big[ (\lambda_1^2-\lambda_2^2)^4 + p^6 (\lambda_1^2+4 \lambda_2^2) - 2 \lambda_2^2 p^4 (\lambda_1^2-3 \lambda_2^2) + p^2 (\lambda_1^2-\lambda_2^2)^2 (\lambda_1^2 + 4 \lambda_2^2) + p^8  \Big] \\ \nonumber
	& \, + \frac{1}{3} \log \lambda_2^2 - \frac{\lambda_2^2-\lambda_1^2+p^2}{6 p^2} - \frac{(\lambda_1^2-\lambda_2^2)^2 + 2 \lambda_2^2 p^2 + p^4}{3 p^4} \\ \nonumber
	& \, + \frac{3 \lambda_2^2 p^2 (\lambda_2^2 - \lambda_1^2 + p^2) - (\lambda_1^2-\lambda_2^2)^3 + p^6}{3 p^6} \log \Big(\frac{\lambda_1}{\lambda_2}\Big) - \frac{2}{9} \,, \\[1ex] \nonumber
	f_3 = & \, \frac{1}{8 p^8 \zeta} D_\text{cut} \Big[ \big((\lambda_1-\lambda_2)^2+p^2\big) \big(\lambda_2^2 - \lambda_1^2 + p^2\big) \big(\lambda_1^4+\lambda_2^4+p^4+2 \lambda_2^2 (p^2 - \lambda_1^2)\big) \big((\lambda_1+\lambda_2)^2+p^2\big) \Big]   \\ \nonumber
	& \, -\frac{1}{8} \log(-\lambda_2^2)+ \frac{1}{4} \log(\lambda_2^2) + \frac{\lambda_1^2-13 \lambda_2^2}{12 p^2} - \frac{\lambda_1^4-8 \lambda_1^2 \lambda_2^2 + 7 \lambda_2^4}{8 p^4} + \frac{(\lambda_1^2-\lambda_2^2)^3}{4 p^6} - \frac{7}{12} \\ \nonumber
	& \, + \frac{\log(-\lambda_1^2)}{8 p^8}  \Big[ \big(\lambda_1^2-\lambda_2^2\big)^4+p^4 \big(6 \lambda_2^4-4 \lambda_1^2 \lambda_2^2\big)+4 \lambda_2^2 p^2 \big(\lambda_1^2-\lambda_2^2\big)^2+4 \lambda_2^2 p^6+p^8 \Big] \\ 
	& \, - \frac{\log(-\lambda_2^2)}{8 p^8} \Big[ \lambda_1^8 + \lambda_2^8 + 4 \lambda_2^6 (p^2 - \lambda_1^2) + 2 \lambda_2^4 \big(3 \lambda_1^4 - 4 \lambda_1^2 p^2+3 p^4\big) + 4 \lambda_2^2(p^2 - \lambda_1^2) \big(\lambda_1^4+p^4\big) \Big]  \,.
\label{eq:res_f_i} \end{align}
where we defined

\begin{align} \label{eq:def_d_cut_zeta}
	D_\text{cut} = \log(\zeta+\lambda_1^2-\lambda_2^2+p^2) -\log(\zeta+\lambda_1^2-\lambda_2^2-p^2) + \log(\zeta-\lambda_1^2+\lambda_2^2 + p^2) - \log(\zeta-\lambda_1^2+\lambda_2^2-p^2) \,,
\end{align}
with $\zeta = \sqrt{\lambda_2^4+\big(\lambda_1^2+p^2\big)^2+2 \lambda_2^2 (p^2 - \lambda_1^2)}$, and
\begin{align} 
	g_0 = & \, \log \lambda_2^2 - \frac{(\lambda_2^2+p^2) \log(-\lambda_2^2)}{p^2} + \Big(\frac{\lambda_2^2}{p^2}+1 \Big) \log(-\lambda_2^2-p^2) - 2 \,, \\ \nonumber
	g_1 = & \, \frac{1}{2 p^4} \Big[ -p^2 (\lambda_2^2+2 p^2) + p^4 \log \lambda_2^2 - \log(-\lambda_2^2) (\lambda_2^2 + p^2)^2 + (\lambda_2^2 + p^2)^2 \log(-\lambda_2^2 - p^2) \Big] \,, \\[1ex] \nonumber
	g_2 = & \, -\frac{1}{18 p^6} \Big[15 \lambda_2^2 p^4+6 \lambda_2^4 p^2-6 p^6 \log(\lambda_2^2) + 6 \log(-\lambda_2^2)(\lambda_2^2 + p^2)^3 - 6(\lambda_2^2 + p^2)^3 \log(-\lambda_2^2 - p^2) + 13 p^6 \Big] \,, \\ \nonumber	
	g_3 = & \, \frac{1}{24 p^8} \Big[-p^2 (6 \lambda_2^6 + 26 \lambda_2^2 p^4 + 21 \lambda_2^4 p^2 + 14 p^6) + 6 p^8 \log(\lambda_2^2) - 6 \log(-\lambda_2^2) (\lambda_2^2 + p^2)^4 + 6 (\lambda_2^2 + p^2)^4 \log(-\lambda_2^2 - p^2) \Big] \,.
\label{eq:res_g_i}\end{align}
\end{widetext}

\subsection{Real frequencies} \label{subsec:appII_real_frequencies}

For a real-time expression of the ghost gluon loop, we need~\labelcref{eq:I_final} at real frequencies $\omega$, i.e. $I(\omega,\lambda_1,\lambda_2) := I(-\text{i}(\omega + \imag 0^+))$. From the definitions of the functions $\alpha_i$~\labelcref{eq:def_alpha_i}, $\beta_i$~\labelcref{eq:def_beta_i}, $g_i$~\labelcref{eq:res_g_i} and $f_i$~\labelcref{eq:res_f_i}, their respective real-time expressions are trivially obtained by the substitution $p \leftrightarrow \imag \omega$. The calculations were performed in \textsc{Wolfram Mathematica 12.1} with the convention $\text{Im} \, \log x  = \pi $ for $x < 0$ for the logarithmic branch cut.

\section{Numerical procedure} \label{app:numerics}
This appendix elaborates on the numerical treatment of the spectral integrals as well as the spectral integrands.

\subsection{Spectral integration and convergence} \label{subsec:spectral_integration}

The spectral integrals of the form
\begin{align}
	\int_{\{\lambda_i\}} \prod_{i} \lambda_i \rho_i(\lambda_i) I_\text{ren}\big(p,\{\lambda_i\}\big) \,,
\end{align}

where $I_\text{ren}$ is the renormalised spectral integrand (comp.~\labelcref{eq:DSEren} or~\labelcref{eq:DSE_realtime}), are evaluated numerically on a logarithmic momentum grid of about 200 grid points with boundary $(p_\text{min},p_\text{max}) = (10^{-4},10^2)$, identically for the Euclidean and Minkowskian axis.  We use a global adaptive integration strategy with default multidimensional symmetric cubature integration rule.  After spectral integration, the diagram is interpolated with splines in the Euclidean and Hermite polynomials in the Minkowski domain, both of order 3. The spectral function is then computed from the interpolants. Note that, a priori, due to~\labelcref{eq:spec_func_def}, the domain of the ghost spectral function is given by the momentum grid. The integration domain of the spectral integral of the ghost spectral parameter has to be bounded by $(p_\text{min},p_\text{max})$, in order to not rely on the extrapolation of the spectral function beyond the grid points. Due to numerical oscillations at the very low end of the grid, we choose $(\lambda_c^\text{min},\lambda_c^\text{max}) = (10^{-3.5},10^2)$. Convergence of the integration result with respect to increase of the integration domain has been explicitly checked.
\begin{figure}[t]
	\centering
	\includegraphics[width=.98\linewidth]{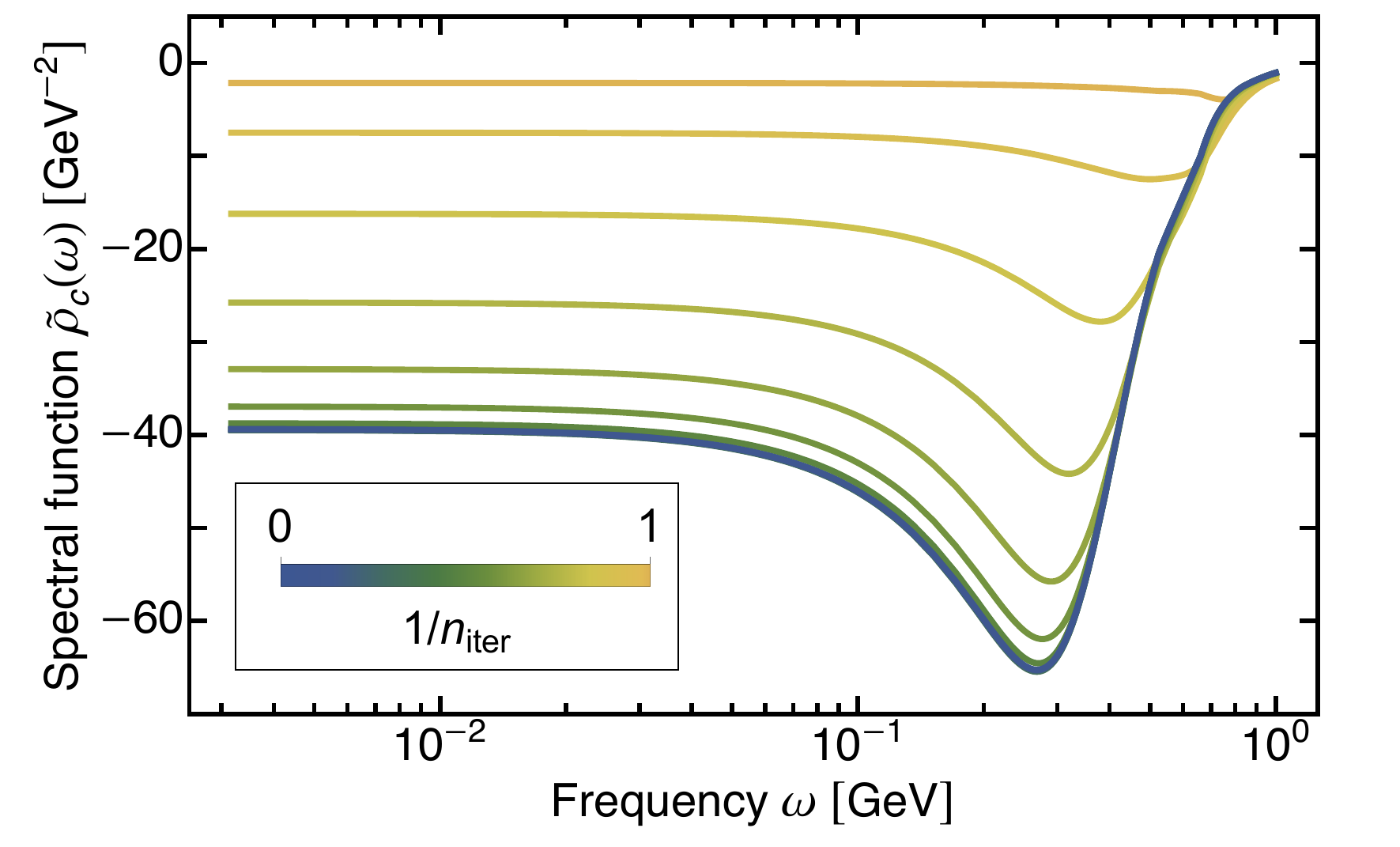}
	\caption{Convergence of an exemplary spectral function through iteration of the DSE. The colour coding indicates the iteration number $n_\text{iter}$. After about 10 iterations, the curves become visually indistinguishable, i.e. the iteration converges.}
	\label{fig:convergence_plot}
\end{figure}

For the gluon spectral integral, the situation is different, as the spectral function is given in an algebraic form from~\cite{Cyrol:2018xeq}. We use the integration boundary $(\lambda_A^\text{min},\lambda_A^\text{max}) = (10^{-4},10^2)$.

\subsection{Spectral integrands} \label{subsec:spectral_integrands}

The numerical performance of the spectral integrations presented in~\Cref{app:loop_mom_int} is sped-up by up to two orders of magnitude by using interpolating functions of the numerical data. The interpolants are constructed by first discretising the integrand inside the three-dimensional $(p,\lambda_c,\lambda_A)$ cuboid defined by $p \in 10^{\{-4,2\}}$, $\lambda_{c/A} \in 10^{\{-4,4\}}$. As for the momentum grid for the spectral integration, we use the same cuboid for the real- and imaginary-time domain. We use 60 grid points in the momentum and 160 grid points in the spectral parameter integration, both with logarithmic grid spacing. For the real-time expressions, we divide into real and imaginary part of the integrands. Both real and imaginary parts of the discretised Minkowski as well as the Euclidean expressions are then interpolated by three dimensional splines inside the cuboid. The resulting interpolating functions are then used in the spectral integration.

\subsection{Convergence of iterative solution} \label{subsec:convergence}

The iteration is described in \sec{sec:SolIt}. It is initiated with a classical spectral function, $	\rho_c^{(0)}(\omega) = \pi \, \delta(\omega^2)$, see also~\labelcref{eq:initial_guess_ghost}. It converges rapidly, see \Cref{fig:convergence_plot}. 

\subsection{Spectral fits} \label{subsec:spectral_fits}

\begin{figure}[t]
	\centering
	\includegraphics[width=.98\linewidth]{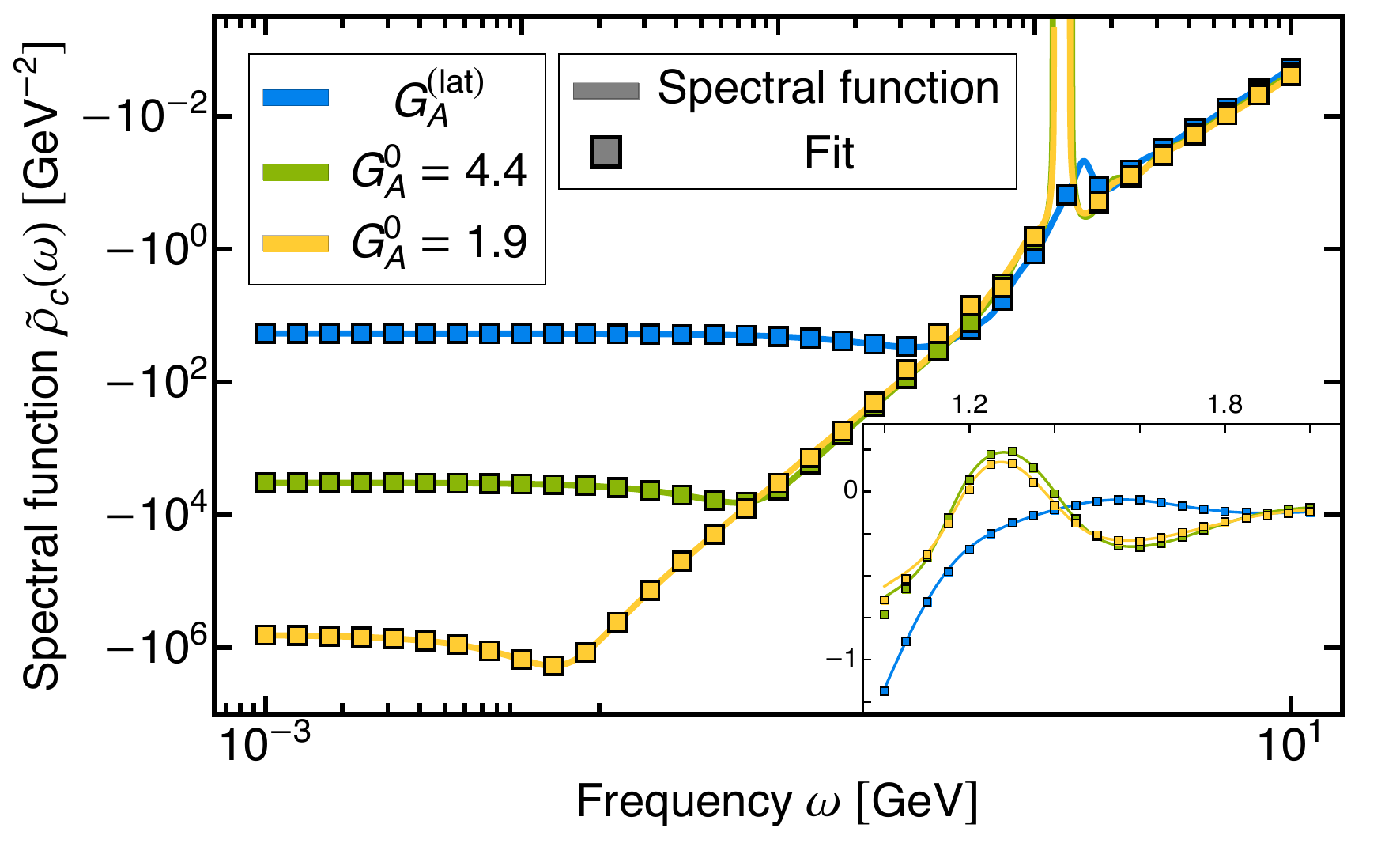}
	\caption{Ghost spectral functions (solid lines) compared to their fits via ansatz~\labelcref{eq:spec_fit}. The best fit parameters are listed in~\Cref{tab:fit_params}. The change of sign around 1.2 GeV is an imprint of the oscillations of the input reconstructed gluon spectral function from~\cite{Cyrol:2018xeq} and is discussed in~\Cref{subsec:spectral_fits}.}
	\label{fig:specs_log_log}
\end{figure}
\begingroup
\definecolor{Gray}{gray}{0.9}
\renewcommand{\arraystretch}{1.1}
\begin{table}[b]
	\centering
	\begin{tabular}{ | c | c | c | c | c | }
		\hline 
		\multicolumn{1}{|>{\columncolor{Gray}[0.7mm][0.7mm]}c|}{\tabularCenterstack{c}{$\bm{{G_A(0)}}$ \\ (lat) \\ 4.4 \\ 1.9}} & \makecell{$\delta_\text{peak}$  \\ 1.89930 \\ 1.61917 \\ 1.13871} & \makecell{$\delta_1$ \\ 10.7596 \\ 3.91085 \\ 20.5415} & \makecell{$\delta_2$ \\ 2.14185 \\ 1.34667 \\ 1.07826} & \makecell{$\delta_3$ \\ 2.15831 \\ 1.88137 \\ 19.8684} \\ \hline
		\makecell{$\gamma_c$ \\ 0.891763 \\ 0.916629 \\ 0.794928} & \makecell{$\hat\Gamma_\text{peak}$ \\ 0.221917 \\ 0.0435632 \\ 0.00641307} & \makecell{$\hat\Gamma_1$ \\ 1.21044 \\ 0.978503 \\ 0.837894 } & \makecell{$\hat\Gamma_2$ \\ 0.408565 \\ 0.245856 \\ 0.151710 } & \makecell{$\hat\Gamma_3$ \\ 0.410932 \\ 0.285878 \\ 0.890227}  \\ \hline 
		\makecell{$Z_\text{UV}$ \\ 0.746555 \\ 0.933136 \\ 0.859889} & \makecell{$\hat M_\text{peak}$ \\ 0.374992 \\ 0.0146429 \\ 0.0134048} & \makecell{$\hat M_1$ \\ 0.589509 \\ 1.12309 \\ 0.654315} & \makecell{$\hat M_2$ \\ 1.59738 \\ 1.26385 \\ 1.26661} & \makecell{$\hat M_3$ \\ 1.59758 \\ 1.55235 \\ 1.55657} \\ \hline
		\makecell{$\hat\rho_0$ \\ 18.4661 \\ 3235.78 \\ 417864} & \makecell{$c_\text{peak}$ \\ 0.656878 \\ 1.90835 \\ 37.4574} & \makecell{$c_1$ \\ 231.204 \\ 1.02717 \\ 0.00294505} & \makecell{$c_2$ \\ -1.1422 \\ -0.0321321 \\ -0.0113776} & \makecell{$c_3$ \\ 1.13776 \\ 0.00133668 \\ 0.00148406}  \\ \hline
		\makecell{$\nu_\text{UV}$ \\ 13.5935 \\ 9.49617 \\ 19.5468} & \makecell{$\nu_\text{IR}$ \\ -9.99927 \\ -49.8529 \\ -195.657} &  \makecell{$\nu_1$ \\ 16.953 \\ 80.9258 \\ 7800.45} & \makecell{$\nu_2$ \\ -28.7442 \\ -13.3501 \\ - 6.45} & \\ \hline
		\makecell{$\hat\Lambda_\text{UV}$ \\ 1.92012 \\ 2.18064 \\ 0.967466} & \makecell{$\hat\Lambda_\text{IR}$ \\ 0.374992 \\ 0.0775041 \\ 0.0816488} &  \makecell{$\hat\Lambda_1$ \\ 0.0998299 \\ 0.0721607 \\ -0.0648268} & \makecell{$\hat\Lambda_2$ \\ 0.759367 \\ 0.762104 \\ 0 } & \\ \hline
	\end{tabular}
	\caption{Best fit parameters for the ansatz~\labelcref{eq:spec_fit} of the spectral functions for the different solutions. As indicated in the top right cell, in each cell the first line contains the fit parameter of the ghost spectral function corresponding to the lattice-like input gluon propagator $G_A^\text{(lat)}$, the second to $G_A(0) = 4.4 \, \mathrm{GeV^{-2}}$ and the last line to the scaling-like $G_A(0) = 1.9 \, \mathrm{GeV^{-2}}$.}
	\label{tab:fit_params}
\end{table}
\endgroup
As discussed in \Cref{sec:results}, we provide a ready-to-use analytic fit formula for the ghost spectral function, see  \eq{eq:spec_fit}. For our best fit  we use $N=3$, the fit parameters for the ghost spectral functions for all input gluon propagators are listed in~\Cref{tab:fit_params}. We show the spectral functions and their respective fits on a log-log scale in~\Cref{fig:specs_log_log}. For $G_A(0) \, [\text{GeV}^{-2}] = 4.4$ and 1.9, the spectral functions feature a change of sign between 1.2 and 1.3 GeV. These wiggles are imprints of the oscillations in the input reconstructed gluon spectral function of~\cite{Cyrol:2018xeq}, and can be understood as numerical artefacts from the reconstruction process. However, in order to match the original Euclidean dressing function $1/Z_c(p)$ in the UV (comp.~\Cref{fig:fit_plot}), it is necessary to keep the respective oscillatory behaviour in the fit.

\vfill

\ 

\eject

\ 



\begin{thebibliography}{79}%
\makeatletter
\providecommand \@ifxundefined [1]{%
 \@ifx{#1\undefined}
}%
\providecommand \@ifnum [1]{%
 \ifnum #1\expandafter \@firstoftwo
 \else \expandafter \@secondoftwo
 \fi
}%
\providecommand \@ifx [1]{%
 \ifx #1\expandafter \@firstoftwo
 \else \expandafter \@secondoftwo
 \fi
}%
\providecommand \natexlab [1]{#1}%
\providecommand \enquote  [1]{``#1''}%
\providecommand \bibnamefont  [1]{#1}%
\providecommand \bibfnamefont [1]{#1}%
\providecommand \citenamefont [1]{#1}%
\providecommand \href@noop [0]{\@secondoftwo}%
\providecommand \href [0]{\begingroup \@sanitize@url \@href}%
\providecommand \@href[1]{\@@startlink{#1}\@@href}%
\providecommand \@@href[1]{\endgroup#1\@@endlink}%
\providecommand \@sanitize@url [0]{\catcode `\\12\catcode `\$12\catcode
  `\&12\catcode `\#12\catcode `\^12\catcode `\_12\catcode `\%12\relax}%
\providecommand \@@startlink[1]{}%
\providecommand \@@endlink[0]{}%
\providecommand \url  [0]{\begingroup\@sanitize@url \@url }%
\providecommand \@url [1]{\endgroup\@href {#1}{\urlprefix }}%
\providecommand \urlprefix  [0]{URL }%
\providecommand \Eprint [0]{\href }%
\providecommand \doibase [0]{http://dx.doi.org/}%
\providecommand \selectlanguage [0]{\@gobble}%
\providecommand \bibinfo  [0]{\@secondoftwo}%
\providecommand \bibfield  [0]{\@secondoftwo}%
\providecommand \translation [1]{[#1]}%
\providecommand \BibitemOpen [0]{}%
\providecommand \bibitemStop [0]{}%
\providecommand \bibitemNoStop [0]{.\EOS\space}%
\providecommand \EOS [0]{\spacefactor3000\relax}%
\providecommand \BibitemShut  [1]{\csname bibitem#1\endcsname}%
\let\auto@bib@innerbib\@empty
\bibitem [{\citenamefont {Horak}\ \emph {et~al.}(2020)\citenamefont {Horak},
  \citenamefont {Pawlowski},\ and\ \citenamefont {Wink}}]{Horak:2020eng}%
  \BibitemOpen
  \bibfield  {author} {\bibinfo {author} {\bibfnamefont {J.}~\bibnamefont
  {Horak}}, \bibinfo {author} {\bibfnamefont {J.~M.}\ \bibnamefont
  {Pawlowski}}, \ and\ \bibinfo {author} {\bibfnamefont {N.}~\bibnamefont
  {Wink}},\ }\href {\doibase 10.1103/PhysRevD.102.125016} {\bibfield  {journal}
  {\bibinfo  {journal} {Phys. Rev. D}\ }\textbf {\bibinfo {volume} {102}},\
  \bibinfo {pages} {125016} (\bibinfo {year} {2020})},\ \Eprint
  {http://arxiv.org/abs/2006.09778} {arXiv:2006.09778 [hep-th]} \BibitemShut
  {NoStop}%
\bibitem [{\citenamefont {Haas}\ \emph {et~al.}(2014)\citenamefont {Haas},
  \citenamefont {Fister},\ and\ \citenamefont {Pawlowski}}]{Haas:2013hpa}%
  \BibitemOpen
  \bibfield  {author} {\bibinfo {author} {\bibfnamefont {M.}~\bibnamefont
  {Haas}}, \bibinfo {author} {\bibfnamefont {L.}~\bibnamefont {Fister}}, \ and\
  \bibinfo {author} {\bibfnamefont {J.~M.}\ \bibnamefont {Pawlowski}},\ }\href
  {\doibase 10.1103/PhysRevD.90.091501} {\bibfield  {journal} {\bibinfo
  {journal} {Phys.Rev.}\ }\textbf {\bibinfo {volume} {D90}},\ \bibinfo {pages}
  {091501} (\bibinfo {year} {2014})},\ \Eprint {http://arxiv.org/abs/1308.4960}
  {arXiv:1308.4960 [hep-ph]} \BibitemShut {NoStop}%
\bibitem [{\citenamefont {Dudal}\ \emph {et~al.}(2014)\citenamefont {Dudal},
  \citenamefont {Oliveira},\ and\ \citenamefont {Silva}}]{Dudal:2013yva}%
  \BibitemOpen
  \bibfield  {author} {\bibinfo {author} {\bibfnamefont {D.}~\bibnamefont
  {Dudal}}, \bibinfo {author} {\bibfnamefont {O.}~\bibnamefont {Oliveira}}, \
  and\ \bibinfo {author} {\bibfnamefont {P.~J.}\ \bibnamefont {Silva}},\ }\href
  {\doibase 10.1103/PhysRevD.89.014010} {\bibfield  {journal} {\bibinfo
  {journal} {Phys. Rev. D}\ }\textbf {\bibinfo {volume} {89}},\ \bibinfo
  {pages} {014010} (\bibinfo {year} {2014})},\ \Eprint
  {http://arxiv.org/abs/1310.4069} {arXiv:1310.4069 [hep-lat]} \BibitemShut
  {NoStop}%
\bibitem [{\citenamefont {Cyrol}\ \emph {et~al.}(2018)\citenamefont {Cyrol},
  \citenamefont {Pawlowski}, \citenamefont {Rothkopf},\ and\ \citenamefont
  {Wink}}]{Cyrol:2018xeq}%
  \BibitemOpen
  \bibfield  {author} {\bibinfo {author} {\bibfnamefont {A.~K.}\ \bibnamefont
  {Cyrol}}, \bibinfo {author} {\bibfnamefont {J.~M.}\ \bibnamefont
  {Pawlowski}}, \bibinfo {author} {\bibfnamefont {A.}~\bibnamefont {Rothkopf}},
  \ and\ \bibinfo {author} {\bibfnamefont {N.}~\bibnamefont {Wink}},\ }\href
  {\doibase 10.21468/SciPostPhys.5.6.065} {\bibfield  {journal} {\bibinfo
  {journal} {SciPost Phys.}\ }\textbf {\bibinfo {volume} {5}},\ \bibinfo
  {pages} {065} (\bibinfo {year} {2018})},\ \Eprint
  {http://arxiv.org/abs/1804.00945} {arXiv:1804.00945 [hep-ph]} \BibitemShut
  {NoStop}%
\bibitem [{\citenamefont {Binosi}\ and\ \citenamefont
  {Tripolt}(2020)}]{Binosi:2019ecz}%
  \BibitemOpen
  \bibfield  {author} {\bibinfo {author} {\bibfnamefont {D.}~\bibnamefont
  {Binosi}}\ and\ \bibinfo {author} {\bibfnamefont {R.-A.}\ \bibnamefont
  {Tripolt}},\ }\href {\doibase 10.1016/j.physletb.2019.135171} {\bibfield
  {journal} {\bibinfo  {journal} {Phys. Lett. B}\ }\textbf {\bibinfo {volume}
  {801}},\ \bibinfo {pages} {135171} (\bibinfo {year} {2020})},\ \Eprint
  {http://arxiv.org/abs/1904.08172} {arXiv:1904.08172 [hep-ph]} \BibitemShut
  {NoStop}%
\bibitem [{\citenamefont {Dudal}\ \emph {et~al.}(2020)\citenamefont {Dudal},
  \citenamefont {Oliveira}, \citenamefont {Roelfs},\ and\ \citenamefont
  {Silva}}]{Dudal:2019gvn}%
  \BibitemOpen
  \bibfield  {author} {\bibinfo {author} {\bibfnamefont {D.}~\bibnamefont
  {Dudal}}, \bibinfo {author} {\bibfnamefont {O.}~\bibnamefont {Oliveira}},
  \bibinfo {author} {\bibfnamefont {M.}~\bibnamefont {Roelfs}}, \ and\ \bibinfo
  {author} {\bibfnamefont {P.}~\bibnamefont {Silva}},\ }\href {\doibase
  10.1016/j.nuclphysb.2019.114912} {\bibfield  {journal} {\bibinfo  {journal}
  {Nucl. Phys. B}\ }\textbf {\bibinfo {volume} {952}},\ \bibinfo {pages}
  {114912} (\bibinfo {year} {2020})},\ \Eprint
  {http://arxiv.org/abs/1901.05348} {arXiv:1901.05348 [hep-lat]} \BibitemShut
  {NoStop}%
\bibitem [{\citenamefont {Siringo}(2016)}]{Siringo:2016jrc}%
  \BibitemOpen
  \bibfield  {author} {\bibinfo {author} {\bibfnamefont {F.}~\bibnamefont
  {Siringo}},\ }\href {\doibase 10.1103/PhysRevD.94.114036} {\bibfield
  {journal} {\bibinfo  {journal} {Phys. Rev. D}\ }\textbf {\bibinfo {volume}
  {94}},\ \bibinfo {pages} {114036} (\bibinfo {year} {2016})},\ \Eprint
  {http://arxiv.org/abs/1605.07357} {arXiv:1605.07357 [hep-ph]} \BibitemShut
  {NoStop}%
\bibitem [{\citenamefont {Hayashi}\ and\ \citenamefont
  {Kondo}(2020)}]{Hayashi:2020few}%
  \BibitemOpen
  \bibfield  {author} {\bibinfo {author} {\bibfnamefont {Y.}~\bibnamefont
  {Hayashi}}\ and\ \bibinfo {author} {\bibfnamefont {K.-I.}\ \bibnamefont
  {Kondo}},\ }\href {\doibase 10.1103/PhysRevD.101.074044} {\bibfield
  {journal} {\bibinfo  {journal} {Phys. Rev. D}\ }\textbf {\bibinfo {volume}
  {101}},\ \bibinfo {pages} {074044} (\bibinfo {year} {2020})},\ \Eprint
  {http://arxiv.org/abs/2001.05987} {arXiv:2001.05987 [hep-th]} \BibitemShut
  {NoStop}%
\bibitem [{\citenamefont {Strauss}\ \emph {et~al.}(2012)\citenamefont
  {Strauss}, \citenamefont {Fischer},\ and\ \citenamefont
  {Kellermann}}]{Strauss:2012dg}%
  \BibitemOpen
  \bibfield  {author} {\bibinfo {author} {\bibfnamefont {S.}~\bibnamefont
  {Strauss}}, \bibinfo {author} {\bibfnamefont {C.~S.}\ \bibnamefont
  {Fischer}}, \ and\ \bibinfo {author} {\bibfnamefont {C.}~\bibnamefont
  {Kellermann}},\ }\href {\doibase 10.1103/PhysRevLett.109.252001} {\bibfield
  {journal} {\bibinfo  {journal} {Phys. Rev. Lett.}\ }\textbf {\bibinfo
  {volume} {109}},\ \bibinfo {pages} {252001} (\bibinfo {year} {2012})},\
  \Eprint {http://arxiv.org/abs/1208.6239} {arXiv:1208.6239 [hep-ph]}
  \BibitemShut {NoStop}%
\bibitem [{\citenamefont {Fischer}\ and\ \citenamefont
  {Huber}(2020)}]{Fischer:2020xnb}%
  \BibitemOpen
  \bibfield  {author} {\bibinfo {author} {\bibfnamefont {C.~S.}\ \bibnamefont
  {Fischer}}\ and\ \bibinfo {author} {\bibfnamefont {M.~Q.}\ \bibnamefont
  {Huber}},\ }\href {\doibase 10.1103/PhysRevD.102.094005} {\bibfield
  {journal} {\bibinfo  {journal} {Phys. Rev. D}\ }\textbf {\bibinfo {volume}
  {102}},\ \bibinfo {pages} {094005} (\bibinfo {year} {2020})},\ \Eprint
  {http://arxiv.org/abs/2007.11505} {arXiv:2007.11505 [hep-ph]} \BibitemShut
  {NoStop}%
\bibitem [{\citenamefont {Sauli}(2020)}]{Sauli:2020dmx}%
  \BibitemOpen
  \bibfield  {author} {\bibinfo {author} {\bibfnamefont {V.}~\bibnamefont
  {Sauli}},\ }\href@noop {} {\  (\bibinfo {year} {2020})},\ \Eprint
  {http://arxiv.org/abs/2011.00536} {arXiv:2011.00536 [hep-lat]} \BibitemShut
  {NoStop}%
\bibitem [{\citenamefont {Pawlowski}(2007)}]{Pawlowski:2005xe}%
  \BibitemOpen
  \bibfield  {author} {\bibinfo {author} {\bibfnamefont {J.~M.}\ \bibnamefont
  {Pawlowski}},\ }\href {\doibase 10.1016/j.aop.2007.01.007} {\bibfield
  {journal} {\bibinfo  {journal} {Annals Phys.}\ }\textbf {\bibinfo {volume}
  {322}},\ \bibinfo {pages} {2831} (\bibinfo {year} {2007})},\ \Eprint
  {http://arxiv.org/abs/hep-th/0512261} {arXiv:hep-th/0512261 [hep-th]}
  \BibitemShut {NoStop}%
\bibitem [{\citenamefont {Gies}(2012)}]{Gies:2006wv}%
  \BibitemOpen
  \bibfield  {author} {\bibinfo {author} {\bibfnamefont {H.}~\bibnamefont
  {Gies}},\ }\href {\doibase 10.1007/978-3-642-27320-9_6} {\bibfield  {journal}
  {\bibinfo  {journal} {Lect.Notes Phys.}\ }\textbf {\bibinfo {volume} {852}},\
  \bibinfo {pages} {287} (\bibinfo {year} {2012})},\ \Eprint
  {http://arxiv.org/abs/hep-ph/0611146} {arXiv:hep-ph/0611146 [hep-ph]}
  \BibitemShut {NoStop}%
\bibitem [{\citenamefont {Rosten}(2012)}]{Rosten:2010vm}%
  \BibitemOpen
  \bibfield  {author} {\bibinfo {author} {\bibfnamefont {O.~J.}\ \bibnamefont
  {Rosten}},\ }\href {\doibase 10.1016/j.physrep.2011.12.003} {\bibfield
  {journal} {\bibinfo  {journal} {Phys. Rept.}\ }\textbf {\bibinfo {volume}
  {511}},\ \bibinfo {pages} {177} (\bibinfo {year} {2012})},\ \Eprint
  {http://arxiv.org/abs/1003.1366} {arXiv:1003.1366 [hep-th]} \BibitemShut
  {NoStop}%
\bibitem [{\citenamefont {Braun}(2012)}]{Braun:2011pp}%
  \BibitemOpen
  \bibfield  {author} {\bibinfo {author} {\bibfnamefont {J.}~\bibnamefont
  {Braun}},\ }\href {\doibase 10.1088/0954-3899/39/3/033001} {\bibfield
  {journal} {\bibinfo  {journal} {J.Phys.}\ }\textbf {\bibinfo {volume}
  {G39}},\ \bibinfo {pages} {033001} (\bibinfo {year} {2012})},\ \Eprint
  {http://arxiv.org/abs/1108.4449} {arXiv:1108.4449 [hep-ph]} \BibitemShut
  {NoStop}%
\bibitem [{\citenamefont {Pawlowski}(2014)}]{Pawlowski:2014aha}%
  \BibitemOpen
  \bibfield  {author} {\bibinfo {author} {\bibfnamefont {J.~M.}\ \bibnamefont
  {Pawlowski}},\ }\href {\doibase 10.1016/j.nuclphysa.2014.09.074} {\bibfield
  {journal} {\bibinfo  {journal} {Nucl.Phys.}\ }\textbf {\bibinfo {volume}
  {A931}},\ \bibinfo {pages} {113} (\bibinfo {year} {2014})}\BibitemShut
  {NoStop}%
\bibitem [{\citenamefont {Dupuis}\ \emph {et~al.}(2021)\citenamefont {Dupuis},
  \citenamefont {Canet}, \citenamefont {Eichhorn}, \citenamefont {Metzner},
  \citenamefont {Pawlowski}, \citenamefont {Tissier},\ and\ \citenamefont
  {Wschebor}}]{Dupuis:2020fhh}%
  \BibitemOpen
  \bibfield  {author} {\bibinfo {author} {\bibfnamefont {N.}~\bibnamefont
  {Dupuis}}, \bibinfo {author} {\bibfnamefont {L.}~\bibnamefont {Canet}},
  \bibinfo {author} {\bibfnamefont {A.}~\bibnamefont {Eichhorn}}, \bibinfo
  {author} {\bibfnamefont {W.}~\bibnamefont {Metzner}}, \bibinfo {author}
  {\bibfnamefont {J.~M.}\ \bibnamefont {Pawlowski}}, \bibinfo {author}
  {\bibfnamefont {M.}~\bibnamefont {Tissier}}, \ and\ \bibinfo {author}
  {\bibfnamefont {N.}~\bibnamefont {Wschebor}},\ }\href {\doibase
  10.1016/j.physrep.2021.01.001} {\bibfield  {journal} {\bibinfo  {journal}
  {Phys. Rept.}\ }\textbf {\bibinfo {volume} {910}},\ \bibinfo {pages} {1}
  (\bibinfo {year} {2021})},\ \Eprint {http://arxiv.org/abs/2006.04853}
  {arXiv:2006.04853 [cond-mat.stat-mech]} \BibitemShut {NoStop}%
\bibitem [{\citenamefont {Roberts}\ and\ \citenamefont
  {Williams}(1994)}]{Roberts:1994dr}%
  \BibitemOpen
  \bibfield  {author} {\bibinfo {author} {\bibfnamefont {C.~D.}\ \bibnamefont
  {Roberts}}\ and\ \bibinfo {author} {\bibfnamefont {A.~G.}\ \bibnamefont
  {Williams}},\ }\href {\doibase 10.1016/0146-6410(94)90049-3} {\bibfield
  {journal} {\bibinfo  {journal} {Prog.Part.Nucl.Phys.}\ }\textbf {\bibinfo
  {volume} {33}},\ \bibinfo {pages} {477} (\bibinfo {year} {1994})},\ \Eprint
  {http://arxiv.org/abs/hep-ph/9403224} {arXiv:hep-ph/9403224 [hep-ph]}
  \BibitemShut {NoStop}%
\bibitem [{\citenamefont {Alkofer}\ and\ \citenamefont {von
  Smekal}(2001)}]{Alkofer:2000wg}%
  \BibitemOpen
  \bibfield  {author} {\bibinfo {author} {\bibfnamefont {R.}~\bibnamefont
  {Alkofer}}\ and\ \bibinfo {author} {\bibfnamefont {L.}~\bibnamefont {von
  Smekal}},\ }\href {\doibase 10.1016/S0370-1573(01)00010-2} {\bibfield
  {journal} {\bibinfo  {journal} {Phys. Rept.}\ }\textbf {\bibinfo {volume}
  {353}},\ \bibinfo {pages} {281} (\bibinfo {year} {2001})},\ \Eprint
  {http://arxiv.org/abs/hep-ph/0007355} {arXiv:hep-ph/0007355} \BibitemShut
  {NoStop}%
\bibitem [{\citenamefont {Maris}\ and\ \citenamefont
  {Roberts}(2003)}]{Maris:2003vk}%
  \BibitemOpen
  \bibfield  {author} {\bibinfo {author} {\bibfnamefont {P.}~\bibnamefont
  {Maris}}\ and\ \bibinfo {author} {\bibfnamefont {C.~D.}\ \bibnamefont
  {Roberts}},\ }\href {\doibase 10.1142/S0218301303001326} {\bibfield
  {journal} {\bibinfo  {journal} {Int.J.Mod.Phys.}\ }\textbf {\bibinfo {volume}
  {E12}},\ \bibinfo {pages} {297} (\bibinfo {year} {2003})},\ \Eprint
  {http://arxiv.org/abs/nucl-th/0301049} {arXiv:nucl-th/0301049 [nucl-th]}
  \BibitemShut {NoStop}%
\bibitem [{\citenamefont {Fischer}(2006)}]{Fischer:2006ub}%
  \BibitemOpen
  \bibfield  {author} {\bibinfo {author} {\bibfnamefont {C.~S.}\ \bibnamefont
  {Fischer}},\ }\href {\doibase 10.1088/0954-3899/32/8/R02} {\bibfield
  {journal} {\bibinfo  {journal} {J.Phys.G}\ }\textbf {\bibinfo {volume}
  {G32}},\ \bibinfo {pages} {R253} (\bibinfo {year} {2006})},\ \Eprint
  {http://arxiv.org/abs/hep-ph/0605173} {arXiv:hep-ph/0605173 [hep-ph]}
  \BibitemShut {NoStop}%
\bibitem [{\citenamefont {Binosi}\ and\ \citenamefont
  {Papavassiliou}(2009)}]{Binosi:2009qm}%
  \BibitemOpen
  \bibfield  {author} {\bibinfo {author} {\bibfnamefont {D.}~\bibnamefont
  {Binosi}}\ and\ \bibinfo {author} {\bibfnamefont {J.}~\bibnamefont
  {Papavassiliou}},\ }\href {\doibase 10.1016/j.physrep.2009.05.001} {\bibfield
   {journal} {\bibinfo  {journal} {Phys.Rept.}\ }\textbf {\bibinfo {volume}
  {479}},\ \bibinfo {pages} {1} (\bibinfo {year} {2009})},\ \Eprint
  {http://arxiv.org/abs/0909.2536} {arXiv:0909.2536 [hep-ph]} \BibitemShut
  {NoStop}%
\bibitem [{\citenamefont {Maas}(2013)}]{Maas:2011se}%
  \BibitemOpen
  \bibfield  {author} {\bibinfo {author} {\bibfnamefont {A.}~\bibnamefont
  {Maas}},\ }\href {\doibase 10.1016/j.physrep.2012.11.002} {\bibfield
  {journal} {\bibinfo  {journal} {Phys.Rept.}\ }\textbf {\bibinfo {volume}
  {524}},\ \bibinfo {pages} {203} (\bibinfo {year} {2013})},\ \Eprint
  {http://arxiv.org/abs/1106.3942} {arXiv:1106.3942 [hep-ph]} \BibitemShut
  {NoStop}%
\bibitem [{\citenamefont {Huber}(2020{\natexlab{a}})}]{Huber:2018ned}%
  \BibitemOpen
  \bibfield  {author} {\bibinfo {author} {\bibfnamefont {M.~Q.}\ \bibnamefont
  {Huber}},\ }\href {\doibase 10.1016/j.physrep.2020.04.004} {\bibfield
  {journal} {\bibinfo  {journal} {Phys. Rept.}\ }\textbf {\bibinfo {volume}
  {879}},\ \bibinfo {pages} {1} (\bibinfo {year} {2020}{\natexlab{a}})},\
  \Eprint {http://arxiv.org/abs/1808.05227} {arXiv:1808.05227 [hep-ph]}
  \BibitemShut {NoStop}%
\bibitem [{\citenamefont {Cloet}\ and\ \citenamefont
  {Roberts}(2014)}]{Cloet:2013jya}%
  \BibitemOpen
  \bibfield  {author} {\bibinfo {author} {\bibfnamefont {I.~C.}\ \bibnamefont
  {Cloet}}\ and\ \bibinfo {author} {\bibfnamefont {C.~D.}\ \bibnamefont
  {Roberts}},\ }\href {\doibase 10.1016/j.ppnp.2014.02.001} {\bibfield
  {journal} {\bibinfo  {journal} {Prog.Part.Nucl.Phys.}\ }\textbf {\bibinfo
  {volume} {77}},\ \bibinfo {pages} {1} (\bibinfo {year} {2014})},\ \Eprint
  {http://arxiv.org/abs/1310.2651} {arXiv:1310.2651 [nucl-th]} \BibitemShut
  {NoStop}%
\bibitem [{\citenamefont {Eichmann}\ \emph {et~al.}(2016)\citenamefont
  {Eichmann}, \citenamefont {Sanchis-Alepuz}, \citenamefont {Williams},
  \citenamefont {Alkofer},\ and\ \citenamefont {Fischer}}]{Eichmann:2016yit}%
  \BibitemOpen
  \bibfield  {author} {\bibinfo {author} {\bibfnamefont {G.}~\bibnamefont
  {Eichmann}}, \bibinfo {author} {\bibfnamefont {H.}~\bibnamefont
  {Sanchis-Alepuz}}, \bibinfo {author} {\bibfnamefont {R.}~\bibnamefont
  {Williams}}, \bibinfo {author} {\bibfnamefont {R.}~\bibnamefont {Alkofer}}, \
  and\ \bibinfo {author} {\bibfnamefont {C.~S.}\ \bibnamefont {Fischer}},\
  }\href {\doibase 10.1016/j.ppnp.2016.07.001} {\bibfield  {journal} {\bibinfo
  {journal} {Prog. Part. Nucl. Phys.}\ }\textbf {\bibinfo {volume} {91}},\
  \bibinfo {pages} {1} (\bibinfo {year} {2016})},\ \Eprint
  {http://arxiv.org/abs/1606.09602} {arXiv:1606.09602 [hep-ph]} \BibitemShut
  {NoStop}%
\bibitem [{\citenamefont {Sanchis-Alepuz}\ and\ \citenamefont
  {Williams}(2018)}]{Sanchis-Alepuz:2017jjd}%
  \BibitemOpen
  \bibfield  {author} {\bibinfo {author} {\bibfnamefont {H.}~\bibnamefont
  {Sanchis-Alepuz}}\ and\ \bibinfo {author} {\bibfnamefont {R.}~\bibnamefont
  {Williams}},\ }\href {\doibase 10.1016/j.cpc.2018.05.020} {\bibfield
  {journal} {\bibinfo  {journal} {Comput. Phys. Commun.}\ }\textbf {\bibinfo
  {volume} {232}},\ \bibinfo {pages} {1} (\bibinfo {year} {2018})},\ \Eprint
  {http://arxiv.org/abs/1710.04903} {arXiv:1710.04903 [hep-ph]} \BibitemShut
  {NoStop}%
\bibitem [{\citenamefont {Kallen}(1952)}]{Kallen:1952zz}%
  \BibitemOpen
  \bibfield  {author} {\bibinfo {author} {\bibfnamefont {G.}~\bibnamefont
  {Kallen}},\ }\href {\doibase 10.1007/978-3-319-00627-7\_90} {\bibfield
  {journal} {\bibinfo  {journal} {Helv. Phys. Acta}\ }\textbf {\bibinfo
  {volume} {25}},\ \bibinfo {pages} {417} (\bibinfo {year} {1952})}\BibitemShut
  {NoStop}%
\bibitem [{\citenamefont {Lehmann}(1954)}]{Lehmann:1954xi}%
  \BibitemOpen
  \bibfield  {author} {\bibinfo {author} {\bibfnamefont {H.}~\bibnamefont
  {Lehmann}},\ }\href {\doibase 10.1007/BF02783624} {\bibfield  {journal}
  {\bibinfo  {journal} {Nuovo Cim.}\ }\textbf {\bibinfo {volume} {11}},\
  \bibinfo {pages} {342} (\bibinfo {year} {1954})}\BibitemShut {NoStop}%
\bibitem [{\citenamefont {Aguilar}\ \emph {et~al.}(2008)\citenamefont
  {Aguilar}, \citenamefont {Binosi},\ and\ \citenamefont
  {Papavassiliou}}]{Aguilar:2008xm}%
  \BibitemOpen
  \bibfield  {author} {\bibinfo {author} {\bibfnamefont {A.~C.}\ \bibnamefont
  {Aguilar}}, \bibinfo {author} {\bibfnamefont {D.}~\bibnamefont {Binosi}}, \
  and\ \bibinfo {author} {\bibfnamefont {J.}~\bibnamefont {Papavassiliou}},\
  }\href {\doibase 10.1103/PhysRevD.78.025010} {\bibfield  {journal} {\bibinfo
  {journal} {Phys. Rev.}\ }\textbf {\bibinfo {volume} {D78}},\ \bibinfo {pages}
  {025010} (\bibinfo {year} {2008})},\ \Eprint {http://arxiv.org/abs/0802.1870}
  {arXiv:0802.1870 [hep-ph]} \BibitemShut {NoStop}%
\bibitem [{\citenamefont {Boucaud}\ \emph {et~al.}(2008)\citenamefont {Boucaud}
  \emph {et~al.}}]{Boucaud:2008ky}%
  \BibitemOpen
  \bibfield  {author} {\bibinfo {author} {\bibfnamefont {P.}~\bibnamefont
  {Boucaud}} \emph {et~al.},\ }\href {\doibase 10.1088/1126-6708/2008/06/099}
  {\bibfield  {journal} {\bibinfo  {journal} {JHEP}\ }\textbf {\bibinfo
  {volume} {06}},\ \bibinfo {pages} {099} (\bibinfo {year} {2008})},\ \Eprint
  {http://arxiv.org/abs/0803.2161} {arXiv:0803.2161 [hep-ph]} \BibitemShut
  {NoStop}%
\bibitem [{\citenamefont {Fischer}\ \emph {et~al.}(2009)\citenamefont
  {Fischer}, \citenamefont {Maas},\ and\ \citenamefont
  {Pawlowski}}]{Fischer:2008uz}%
  \BibitemOpen
  \bibfield  {author} {\bibinfo {author} {\bibfnamefont {C.~S.}\ \bibnamefont
  {Fischer}}, \bibinfo {author} {\bibfnamefont {A.}~\bibnamefont {Maas}}, \
  and\ \bibinfo {author} {\bibfnamefont {J.~M.}\ \bibnamefont {Pawlowski}},\
  }\href {\doibase 10.1016/j.aop.2009.07.009} {\bibfield  {journal} {\bibinfo
  {journal} {Annals Phys.}\ }\textbf {\bibinfo {volume} {324}},\ \bibinfo
  {pages} {2408} (\bibinfo {year} {2009})},\ \Eprint
  {http://arxiv.org/abs/0810.1987} {arXiv:0810.1987 [hep-ph]} \BibitemShut
  {NoStop}%
\bibitem [{\citenamefont {Cyrol}\ \emph {et~al.}(2016)\citenamefont {Cyrol},
  \citenamefont {Fister}, \citenamefont {Mitter}, \citenamefont {Pawlowski},\
  and\ \citenamefont {Strodthoff}}]{Cyrol:2016tym}%
  \BibitemOpen
  \bibfield  {author} {\bibinfo {author} {\bibfnamefont {A.~K.}\ \bibnamefont
  {Cyrol}}, \bibinfo {author} {\bibfnamefont {L.}~\bibnamefont {Fister}},
  \bibinfo {author} {\bibfnamefont {M.}~\bibnamefont {Mitter}}, \bibinfo
  {author} {\bibfnamefont {J.~M.}\ \bibnamefont {Pawlowski}}, \ and\ \bibinfo
  {author} {\bibfnamefont {N.}~\bibnamefont {Strodthoff}},\ }\href {\doibase
  10.1103/PhysRevD.94.054005} {\bibfield  {journal} {\bibinfo  {journal} {Phys.
  Rev.}\ }\textbf {\bibinfo {volume} {D94}},\ \bibinfo {pages} {054005}
  (\bibinfo {year} {2016})},\ \Eprint {http://arxiv.org/abs/1605.01856}
  {arXiv:1605.01856 [hep-ph]} \BibitemShut {NoStop}%
\bibitem [{\citenamefont {Lowdon}(2018{\natexlab{a}})}]{Lowdon:2017gpp}%
  \BibitemOpen
  \bibfield  {author} {\bibinfo {author} {\bibfnamefont {P.}~\bibnamefont
  {Lowdon}},\ }\href {\doibase 10.1016/j.nuclphysb.2018.08.012} {\bibfield
  {journal} {\bibinfo  {journal} {Nucl. Phys. B}\ }\textbf {\bibinfo {volume}
  {935}},\ \bibinfo {pages} {242} (\bibinfo {year} {2018}{\natexlab{a}})},\
  \Eprint {http://arxiv.org/abs/1711.07569} {arXiv:1711.07569 [hep-th]}
  \BibitemShut {NoStop}%
\bibitem [{\citenamefont {Oehme}\ and\ \citenamefont
  {Zimmermann}(1980)}]{Oehme:1979ai}%
  \BibitemOpen
  \bibfield  {author} {\bibinfo {author} {\bibfnamefont {R.}~\bibnamefont
  {Oehme}}\ and\ \bibinfo {author} {\bibfnamefont {W.}~\bibnamefont
  {Zimmermann}},\ }\href {\doibase 10.1103/PhysRevD.21.471} {\bibfield
  {journal} {\bibinfo  {journal} {Phys. Rev.}\ }\textbf {\bibinfo {volume}
  {D21}},\ \bibinfo {pages} {471} (\bibinfo {year} {1980})}\BibitemShut
  {NoStop}%
\bibitem [{\citenamefont {Oehme}(1990)}]{Oehme:1990kd}%
  \BibitemOpen
  \bibfield  {author} {\bibinfo {author} {\bibfnamefont {R.}~\bibnamefont
  {Oehme}},\ }\href {\doibase 10.1016/0370-2693(90)90499-V} {\bibfield
  {journal} {\bibinfo  {journal} {Phys. Lett. B}\ }\textbf {\bibinfo {volume}
  {252}},\ \bibinfo {pages} {641} (\bibinfo {year} {1990})}\BibitemShut
  {NoStop}%
\bibitem [{\citenamefont {Bonanno}\ \emph {et~al.}(2021)\citenamefont
  {Bonanno}, \citenamefont {Denz}, \citenamefont {Pawlowski},\ and\
  \citenamefont {Reichert}}]{Bonanno:2021squ}%
  \BibitemOpen
  \bibfield  {author} {\bibinfo {author} {\bibfnamefont {A.}~\bibnamefont
  {Bonanno}}, \bibinfo {author} {\bibfnamefont {T.}~\bibnamefont {Denz}},
  \bibinfo {author} {\bibfnamefont {J.~M.}\ \bibnamefont {Pawlowski}}, \ and\
  \bibinfo {author} {\bibfnamefont {M.}~\bibnamefont {Reichert}},\ }\href@noop
  {} {\  (\bibinfo {year} {2021})},\ \Eprint {http://arxiv.org/abs/2102.02217}
  {arXiv:2102.02217 [hep-th]} \BibitemShut {NoStop}%
\bibitem [{\citenamefont {Dudal}\ \emph {et~al.}(2008)\citenamefont {Dudal},
  \citenamefont {Gracey}, \citenamefont {Sorella}, \citenamefont
  {Vandersickel},\ and\ \citenamefont {Verschelde}}]{Dudal:2008sp}%
  \BibitemOpen
  \bibfield  {author} {\bibinfo {author} {\bibfnamefont {D.}~\bibnamefont
  {Dudal}}, \bibinfo {author} {\bibfnamefont {J.~A.}\ \bibnamefont {Gracey}},
  \bibinfo {author} {\bibfnamefont {S.~P.}\ \bibnamefont {Sorella}}, \bibinfo
  {author} {\bibfnamefont {N.}~\bibnamefont {Vandersickel}}, \ and\ \bibinfo
  {author} {\bibfnamefont {H.}~\bibnamefont {Verschelde}},\ }\href {\doibase
  10.1103/PhysRevD.78.065047} {\bibfield  {journal} {\bibinfo  {journal}
  {Phys.Rev.}\ }\textbf {\bibinfo {volume} {D78}},\ \bibinfo {pages} {065047}
  (\bibinfo {year} {2008})},\ \Eprint {http://arxiv.org/abs/0806.4348}
  {arXiv:0806.4348 [hep-th]} \BibitemShut {NoStop}%
\bibitem [{\citenamefont {Sorella}(2011)}]{Sorella:2010it}%
  \BibitemOpen
  \bibfield  {author} {\bibinfo {author} {\bibfnamefont {S.~P.}\ \bibnamefont
  {Sorella}},\ }\href {\doibase 10.1088/1751-8113/44/13/135403} {\bibfield
  {journal} {\bibinfo  {journal} {J.Phys.A}\ }\textbf {\bibinfo {volume}
  {A44}},\ \bibinfo {pages} {135403} (\bibinfo {year} {2011})},\ \Eprint
  {http://arxiv.org/abs/1006.4500} {arXiv:1006.4500 [hep-th]} \BibitemShut
  {NoStop}%
\bibitem [{\citenamefont {Lowdon}(2017)}]{Lowdon:2017uqe}%
  \BibitemOpen
  \bibfield  {author} {\bibinfo {author} {\bibfnamefont {P.}~\bibnamefont
  {Lowdon}},\ }\href {\doibase 10.1103/PhysRevD.96.065013} {\bibfield
  {journal} {\bibinfo  {journal} {Phys. Rev. D}\ }\textbf {\bibinfo {volume}
  {96}},\ \bibinfo {pages} {065013} (\bibinfo {year} {2017})},\ \Eprint
  {http://arxiv.org/abs/1702.02954} {arXiv:1702.02954 [hep-th]} \BibitemShut
  {NoStop}%
\bibitem [{\citenamefont {Hayashi}\ and\ \citenamefont
  {Kondo}(2019)}]{Hayashi:2018giz}%
  \BibitemOpen
  \bibfield  {author} {\bibinfo {author} {\bibfnamefont {Y.}~\bibnamefont
  {Hayashi}}\ and\ \bibinfo {author} {\bibfnamefont {K.-I.}\ \bibnamefont
  {Kondo}},\ }\href {\doibase 10.1103/PhysRevD.99.074001} {\bibfield  {journal}
  {\bibinfo  {journal} {Phys. Rev. D}\ }\textbf {\bibinfo {volume} {99}},\
  \bibinfo {pages} {074001} (\bibinfo {year} {2019})},\ \Eprint
  {http://arxiv.org/abs/1812.03116} {arXiv:1812.03116 [hep-th]} \BibitemShut
  {NoStop}%
\bibitem [{\citenamefont {Lowdon}(2018{\natexlab{b}})}]{Lowdon:2018uzf}%
  \BibitemOpen
  \bibfield  {author} {\bibinfo {author} {\bibfnamefont {P.}~\bibnamefont
  {Lowdon}},\ }\href {\doibase 10.22323/1.336.0050} {\bibfield  {journal}
  {\bibinfo  {journal} {PoS}\ }\textbf {\bibinfo {volume} {Confinement2018}},\
  \bibinfo {pages} {050} (\bibinfo {year} {2018}{\natexlab{b}})},\ \Eprint
  {http://arxiv.org/abs/1811.03037} {arXiv:1811.03037 [hep-th]} \BibitemShut
  {NoStop}%
\bibitem [{\citenamefont {Li}\ \emph {et~al.}(2020)\citenamefont {Li},
  \citenamefont {Lowdon}, \citenamefont {Oliveira},\ and\ \citenamefont
  {Silva}}]{Li:2019hyv}%
  \BibitemOpen
  \bibfield  {author} {\bibinfo {author} {\bibfnamefont {S.~W.}\ \bibnamefont
  {Li}}, \bibinfo {author} {\bibfnamefont {P.}~\bibnamefont {Lowdon}}, \bibinfo
  {author} {\bibfnamefont {O.}~\bibnamefont {Oliveira}}, \ and\ \bibinfo
  {author} {\bibfnamefont {P.~J.}\ \bibnamefont {Silva}},\ }\href {\doibase
  10.1016/j.physletb.2020.135329} {\bibfield  {journal} {\bibinfo  {journal}
  {Phys. Lett. B}\ }\textbf {\bibinfo {volume} {803}},\ \bibinfo {pages}
  {135329} (\bibinfo {year} {2020})},\ \Eprint
  {http://arxiv.org/abs/1907.10073} {arXiv:1907.10073 [hep-th]} \BibitemShut
  {NoStop}%
\bibitem [{\citenamefont {Hayashi}\ and\ \citenamefont
  {Kondo}(2021)}]{Hayashi:2021nnj}%
  \BibitemOpen
  \bibfield  {author} {\bibinfo {author} {\bibfnamefont {Y.}~\bibnamefont
  {Hayashi}}\ and\ \bibinfo {author} {\bibfnamefont {K.-I.}\ \bibnamefont
  {Kondo}},\ }\href {\doibase 10.1103/PhysRevD.103.L111504} {\bibfield
  {journal} {\bibinfo  {journal} {Phys. Rev. D}\ }\textbf {\bibinfo {volume}
  {103}},\ \bibinfo {pages} {L111504} (\bibinfo {year} {2021})},\ \Eprint
  {http://arxiv.org/abs/2103.14322} {arXiv:2103.14322 [hep-th]} \BibitemShut
  {NoStop}%
\bibitem [{\citenamefont {Kondo}\ \emph {et~al.}(2019)\citenamefont {Kondo},
  \citenamefont {Hayashi}, \citenamefont {Matsudo}, \citenamefont {Suda},\ and\
  \citenamefont {Watanabe}}]{Kondo:2019ywt}%
  \BibitemOpen
  \bibfield  {author} {\bibinfo {author} {\bibfnamefont {K.-I.}\ \bibnamefont
  {Kondo}}, \bibinfo {author} {\bibfnamefont {Y.}~\bibnamefont {Hayashi}},
  \bibinfo {author} {\bibfnamefont {R.}~\bibnamefont {Matsudo}}, \bibinfo
  {author} {\bibfnamefont {Y.}~\bibnamefont {Suda}}, \ and\ \bibinfo {author}
  {\bibfnamefont {M.}~\bibnamefont {Watanabe}},\ }\href {\doibase
  10.22323/1.374.0053} {\bibfield  {journal} {\bibinfo  {journal} {PoS}\
  }\textbf {\bibinfo {volume} {LC2019}},\ \bibinfo {pages} {053} (\bibinfo
  {year} {2019})},\ \Eprint {http://arxiv.org/abs/1912.06261} {arXiv:1912.06261
  [hep-th]} \BibitemShut {NoStop}%
\bibitem [{\citenamefont {Kondo}\ \emph {et~al.}(2020)\citenamefont {Kondo},
  \citenamefont {Watanabe}, \citenamefont {Hayashi}, \citenamefont {Matsudo},\
  and\ \citenamefont {Suda}}]{Kondo:2019rpa}%
  \BibitemOpen
  \bibfield  {author} {\bibinfo {author} {\bibfnamefont {K.-I.}\ \bibnamefont
  {Kondo}}, \bibinfo {author} {\bibfnamefont {M.}~\bibnamefont {Watanabe}},
  \bibinfo {author} {\bibfnamefont {Y.}~\bibnamefont {Hayashi}}, \bibinfo
  {author} {\bibfnamefont {R.}~\bibnamefont {Matsudo}}, \ and\ \bibinfo
  {author} {\bibfnamefont {Y.}~\bibnamefont {Suda}},\ }\href {\doibase
  10.1140/epjc/s10052-020-7632-4} {\bibfield  {journal} {\bibinfo  {journal}
  {Eur. Phys. J. C}\ }\textbf {\bibinfo {volume} {80}},\ \bibinfo {pages} {84}
  (\bibinfo {year} {2020})},\ \Eprint {http://arxiv.org/abs/1902.08894}
  {arXiv:1902.08894 [hep-th]} \BibitemShut {NoStop}%
\bibitem [{\citenamefont {Schleifenbaum}\ \emph {et~al.}(2005)\citenamefont
  {Schleifenbaum}, \citenamefont {Maas}, \citenamefont {Wambach},\ and\
  \citenamefont {Alkofer}}]{Schleifenbaum:2004id}%
  \BibitemOpen
  \bibfield  {author} {\bibinfo {author} {\bibfnamefont {W.}~\bibnamefont
  {Schleifenbaum}}, \bibinfo {author} {\bibfnamefont {A.}~\bibnamefont {Maas}},
  \bibinfo {author} {\bibfnamefont {J.}~\bibnamefont {Wambach}}, \ and\
  \bibinfo {author} {\bibfnamefont {R.}~\bibnamefont {Alkofer}},\ }\href
  {\doibase 10.1103/PhysRevD.72.014017} {\bibfield  {journal} {\bibinfo
  {journal} {Phys. Rev.}\ }\textbf {\bibinfo {volume} {D72}},\ \bibinfo {pages}
  {014017} (\bibinfo {year} {2005})},\ \Eprint
  {http://arxiv.org/abs/hep-ph/0411052} {arXiv:hep-ph/0411052} \BibitemShut
  {NoStop}%
\bibitem [{\citenamefont {Sternbeck}(2006)}]{Sternbeck:2006rd}%
  \BibitemOpen
  \bibfield  {author} {\bibinfo {author} {\bibfnamefont {A.}~\bibnamefont
  {Sternbeck}},\ }\emph {\bibinfo {title} {{The infrared behavior of lattice
  QCD Green's functions}}},\ \href@noop {} {Ph.D. thesis},\ \bibinfo  {school}
  {Humboldt-University Berlin} (\bibinfo {year} {2006}),\ \Eprint
  {http://arxiv.org/abs/hep-lat/0609016} {arXiv:hep-lat/0609016} \BibitemShut
  {NoStop}%
\bibitem [{\citenamefont {Ilgenfritz}\ \emph {et~al.}(2007)\citenamefont
  {Ilgenfritz}, \citenamefont {M{\"u}ller-Preussker}, \citenamefont
  {Sternbeck}, \citenamefont {Schiller},\ and\ \citenamefont
  {Bogolubsky}}]{Ilgenfritz:2006he}%
  \BibitemOpen
  \bibfield  {author} {\bibinfo {author} {\bibfnamefont {E.~M.}\ \bibnamefont
  {Ilgenfritz}}, \bibinfo {author} {\bibfnamefont {M.}~\bibnamefont
  {M{\"u}ller-Preussker}}, \bibinfo {author} {\bibfnamefont {A.}~\bibnamefont
  {Sternbeck}}, \bibinfo {author} {\bibfnamefont {A.}~\bibnamefont {Schiller}},
  \ and\ \bibinfo {author} {\bibfnamefont {I.~L.}\ \bibnamefont {Bogolubsky}},\
  }\href@noop {} {\bibfield  {journal} {\bibinfo  {journal} {Braz. J. Phys.}\
  }\textbf {\bibinfo {volume} {37}},\ \bibinfo {pages} {193} (\bibinfo {year}
  {2007})},\ \Eprint {http://arxiv.org/abs/hep-lat/0609043}
  {arXiv:hep-lat/0609043} \BibitemShut {NoStop}%
\bibitem [{\citenamefont {Boucaud}\ \emph {et~al.}(2011)\citenamefont
  {Boucaud}, \citenamefont {Dudal}, \citenamefont {Leroy}, \citenamefont
  {Pene},\ and\ \citenamefont {Rodriguez-Quintero}}]{Boucaud:2011eh}%
  \BibitemOpen
  \bibfield  {author} {\bibinfo {author} {\bibfnamefont {P.}~\bibnamefont
  {Boucaud}}, \bibinfo {author} {\bibfnamefont {D.}~\bibnamefont {Dudal}},
  \bibinfo {author} {\bibfnamefont {J.~P.}\ \bibnamefont {Leroy}}, \bibinfo
  {author} {\bibfnamefont {O.}~\bibnamefont {Pene}}, \ and\ \bibinfo {author}
  {\bibfnamefont {J.}~\bibnamefont {Rodriguez-Quintero}},\ }\href {\doibase
  10.1007/JHEP12(2011)018} {\bibfield  {journal} {\bibinfo  {journal} {JHEP}\
  }\textbf {\bibinfo {volume} {12}},\ \bibinfo {pages} {018} (\bibinfo {year}
  {2011})},\ \Eprint {http://arxiv.org/abs/1109.3803} {arXiv:1109.3803
  [hep-ph]} \BibitemShut {NoStop}%
\bibitem [{\citenamefont {Dudal}\ \emph {et~al.}(2012)\citenamefont {Dudal},
  \citenamefont {Oliveira},\ and\ \citenamefont
  {Rodriguez-Quintero}}]{Dudal:2012zx}%
  \BibitemOpen
  \bibfield  {author} {\bibinfo {author} {\bibfnamefont {D.}~\bibnamefont
  {Dudal}}, \bibinfo {author} {\bibfnamefont {O.}~\bibnamefont {Oliveira}}, \
  and\ \bibinfo {author} {\bibfnamefont {J.}~\bibnamefont
  {Rodriguez-Quintero}},\ }\href {\doibase 10.1103/PhysRevD.86.105005,
  10.1103/PhysRevD.86.109902} {\bibfield  {journal} {\bibinfo  {journal}
  {Phys.Rev.}\ }\textbf {\bibinfo {volume} {D86}},\ \bibinfo {pages} {105005}
  (\bibinfo {year} {2012})},\ \Eprint {http://arxiv.org/abs/1207.5118}
  {arXiv:1207.5118 [hep-ph]} \BibitemShut {NoStop}%
\bibitem [{\citenamefont {Aguilar}\ \emph {et~al.}(2013)\citenamefont
  {Aguilar}, \citenamefont {Ib{\'a}{\~n}ez},\ and\ \citenamefont
  {Papavassiliou}}]{Aguilar:2013xqa}%
  \BibitemOpen
  \bibfield  {author} {\bibinfo {author} {\bibfnamefont {A.~C.}\ \bibnamefont
  {Aguilar}}, \bibinfo {author} {\bibfnamefont {D.}~\bibnamefont
  {Ib{\'a}{\~n}ez}}, \ and\ \bibinfo {author} {\bibfnamefont {J.}~\bibnamefont
  {Papavassiliou}},\ }\href {\doibase 10.1103/PhysRevD.87.114020} {\bibfield
  {journal} {\bibinfo  {journal} {Phys. Rev.}\ }\textbf {\bibinfo {volume}
  {D87}},\ \bibinfo {pages} {114020} (\bibinfo {year} {2013})},\ \Eprint
  {http://arxiv.org/abs/1303.3609} {arXiv:1303.3609 [hep-ph]} \BibitemShut
  {NoStop}%
\bibitem [{\citenamefont {Huber}(2020{\natexlab{b}})}]{Huber:2020keu}%
  \BibitemOpen
  \bibfield  {author} {\bibinfo {author} {\bibfnamefont {M.~Q.}\ \bibnamefont
  {Huber}},\ }\href {\doibase 10.1103/PhysRevD.101.114009} {\bibfield
  {journal} {\bibinfo  {journal} {Phys. Rev. D}\ }\textbf {\bibinfo {volume}
  {101}},\ \bibinfo {pages} {114009} (\bibinfo {year} {2020}{\natexlab{b}})},\
  \Eprint {http://arxiv.org/abs/2003.13703} {arXiv:2003.13703 [hep-ph]}
  \BibitemShut {NoStop}%
\bibitem [{\citenamefont {Barrios}\ \emph {et~al.}(2020)\citenamefont
  {Barrios}, \citenamefont {Pel\'aez}, \citenamefont {Reinosa},\ and\
  \citenamefont {Wschebor}}]{Barrios:2020ubx}%
  \BibitemOpen
  \bibfield  {author} {\bibinfo {author} {\bibfnamefont {N.}~\bibnamefont
  {Barrios}}, \bibinfo {author} {\bibfnamefont {M.}~\bibnamefont {Pel\'aez}},
  \bibinfo {author} {\bibfnamefont {U.}~\bibnamefont {Reinosa}}, \ and\
  \bibinfo {author} {\bibfnamefont {N.}~\bibnamefont {Wschebor}},\ }\href
  {\doibase 10.1103/PhysRevD.102.114016} {\bibfield  {journal} {\bibinfo
  {journal} {Phys. Rev. D}\ }\textbf {\bibinfo {volume} {102}},\ \bibinfo
  {pages} {114016} (\bibinfo {year} {2020})},\ \Eprint
  {http://arxiv.org/abs/2009.00875} {arXiv:2009.00875 [hep-th]} \BibitemShut
  {NoStop}%
\bibitem [{\citenamefont {Evans}(1992)}]{Evans:1991ky}%
  \BibitemOpen
  \bibfield  {author} {\bibinfo {author} {\bibfnamefont {T.~S.}\ \bibnamefont
  {Evans}},\ }\href {\doibase 10.1016/0550-3213(92)90357-H} {\bibfield
  {journal} {\bibinfo  {journal} {Nucl. Phys. B}\ }\textbf {\bibinfo {volume}
  {374}},\ \bibinfo {pages} {340} (\bibinfo {year} {1992})}\BibitemShut
  {NoStop}%
\bibitem [{\citenamefont {Aurenche}\ and\ \citenamefont
  {Becherrawy}(1992)}]{Aurenche:1991hi}%
  \BibitemOpen
  \bibfield  {author} {\bibinfo {author} {\bibfnamefont {P.}~\bibnamefont
  {Aurenche}}\ and\ \bibinfo {author} {\bibfnamefont {T.}~\bibnamefont
  {Becherrawy}},\ }\href {\doibase 10.1016/0550-3213(92)90597-5} {\bibfield
  {journal} {\bibinfo  {journal} {Nucl.Phys.}\ }\textbf {\bibinfo {volume}
  {B379}},\ \bibinfo {pages} {259} (\bibinfo {year} {1992})}\BibitemShut
  {NoStop}%
\bibitem [{\citenamefont {Baier}\ and\ \citenamefont
  {Ni\'egawa}(1994)}]{PhysRevD.49.4107}%
  \BibitemOpen
  \bibfield  {author} {\bibinfo {author} {\bibfnamefont {R.}~\bibnamefont
  {Baier}}\ and\ \bibinfo {author} {\bibfnamefont {A.}~\bibnamefont
  {Ni\'egawa}},\ }\href {\doibase 10.1103/PhysRevD.49.4107} {\bibfield
  {journal} {\bibinfo  {journal} {Phys. Rev. D}\ }\textbf {\bibinfo {volume}
  {49}},\ \bibinfo {pages} {4107} (\bibinfo {year} {1994})}\BibitemShut
  {NoStop}%
\bibitem [{\citenamefont {Guerin}(1994)}]{Guerin_1994}%
  \BibitemOpen
  \bibfield  {author} {\bibinfo {author} {\bibfnamefont {F.}~\bibnamefont
  {Guerin}},\ }\href {\doibase 10.1016/0550-3213(94)90603-3} {\bibfield
  {journal} {\bibinfo  {journal} {Nuclear Physics B}\ }\textbf {\bibinfo
  {volume} {432}},\ \bibinfo {pages} {281–311} (\bibinfo {year}
  {1994})}\BibitemShut {NoStop}%
\bibitem [{\citenamefont {Wink}(2020)}]{Wink:2020tnu}%
  \BibitemOpen
  \bibfield  {author} {\bibinfo {author} {\bibfnamefont {N.}~\bibnamefont
  {Wink}},\ }\emph {\bibinfo {title} {{Towards the spectral properties and
  phase structure of QCD.}}},\ \href {\doibase 10.11588/heidok.00028503} {Ph.D.
  thesis},\ \bibinfo  {school} {U. Heidelberg, ITP} (\bibinfo {year}
  {2020})\BibitemShut {NoStop}%
\bibitem [{\citenamefont {Carrington}\ and\ \citenamefont
  {Heinz}(1998)}]{Carrington_1998}%
  \BibitemOpen
  \bibfield  {author} {\bibinfo {author} {\bibfnamefont {M.}~\bibnamefont
  {Carrington}}\ and\ \bibinfo {author} {\bibfnamefont {U.}~\bibnamefont
  {Heinz}},\ }\href {\doibase 10.1007/s100520050110} {\bibfield  {journal}
  {\bibinfo  {journal} {The European Physical Journal C}\ }\textbf {\bibinfo
  {volume} {1}},\ \bibinfo {pages} {619–625} (\bibinfo {year}
  {1998})}\BibitemShut {NoStop}%
\bibitem [{\citenamefont {Bros}\ and\ \citenamefont
  {Buchholz}(1996)}]{Bros:1996mw}%
  \BibitemOpen
  \bibfield  {author} {\bibinfo {author} {\bibfnamefont {J.}~\bibnamefont
  {Bros}}\ and\ \bibinfo {author} {\bibfnamefont {D.}~\bibnamefont
  {Buchholz}},\ }\href@noop {} {\bibfield  {journal} {\bibinfo  {journal} {Ann.
  Inst. H. Poincare Phys. Theor.}\ }\textbf {\bibinfo {volume} {64}},\ \bibinfo
  {pages} {495} (\bibinfo {year} {1996})},\ \Eprint
  {http://arxiv.org/abs/hep-th/9606046} {arXiv:hep-th/9606046 [hep-th]}
  \BibitemShut {NoStop}%
\bibitem [{\citenamefont {Defu}\ and\ \citenamefont {Heinz}(1998)}]{Defu_1998}%
  \BibitemOpen
  \bibfield  {author} {\bibinfo {author} {\bibfnamefont {H.}~\bibnamefont
  {Defu}}\ and\ \bibinfo {author} {\bibfnamefont {U.}~\bibnamefont {Heinz}},\
  }\href {\doibase 10.1007/s100529800748} {\bibfield  {journal} {\bibinfo
  {journal} {The European Physical Journal C}\ }\textbf {\bibinfo {volume}
  {4}},\ \bibinfo {pages} {129–137} (\bibinfo {year} {1998})}\BibitemShut
  {NoStop}%
\bibitem [{\citenamefont {Weldon}(1998)}]{Weldon_1998}%
  \BibitemOpen
  \bibfield  {author} {\bibinfo {author} {\bibfnamefont {H.}~\bibnamefont
  {Weldon}},\ }\href {\doibase 10.1016/s0550-3213(98)00544-6} {\bibfield
  {journal} {\bibinfo  {journal} {Nuclear Physics B}\ }\textbf {\bibinfo
  {volume} {534}},\ \bibinfo {pages} {467–490} (\bibinfo {year}
  {1998})}\BibitemShut {NoStop}%
\bibitem [{\citenamefont {Hou}\ \emph {et~al.}(1998)\citenamefont {Hou},
  \citenamefont {Wang},\ and\ \citenamefont {Heinz}}]{Hou_1998}%
  \BibitemOpen
  \bibfield  {author} {\bibinfo {author} {\bibfnamefont {D.}~\bibnamefont
  {Hou}}, \bibinfo {author} {\bibfnamefont {E.}~\bibnamefont {Wang}}, \ and\
  \bibinfo {author} {\bibfnamefont {U.}~\bibnamefont {Heinz}},\ }\href
  {\doibase 10.1088/0954-3899/24/10/004} {\bibfield  {journal} {\bibinfo
  {journal} {Journal of Physics G: Nuclear and Particle Physics}\ }\textbf
  {\bibinfo {volume} {24}},\ \bibinfo {pages} {1861–1868} (\bibinfo {year}
  {1998})}\BibitemShut {NoStop}%
\bibitem [{\citenamefont {Hou}\ \emph {et~al.}(2000)\citenamefont {Hou},
  \citenamefont {Carrington}, \citenamefont {Kobes},\ and\ \citenamefont
  {Heinz}}]{Hou:1999zv}%
  \BibitemOpen
  \bibfield  {author} {\bibinfo {author} {\bibfnamefont {D.-f.}\ \bibnamefont
  {Hou}}, \bibinfo {author} {\bibfnamefont {M.~E.}\ \bibnamefont {Carrington}},
  \bibinfo {author} {\bibfnamefont {R.}~\bibnamefont {Kobes}}, \ and\ \bibinfo
  {author} {\bibfnamefont {U.~W.}\ \bibnamefont {Heinz}},\ }\href {\doibase
  10.1103/PhysRevD.67.049902} {\bibfield  {journal} {\bibinfo  {journal} {Phys.
  Rev. D}\ }\textbf {\bibinfo {volume} {61}},\ \bibinfo {pages} {085013}
  (\bibinfo {year} {2000})},\ \bibinfo {note} {[Erratum: Phys.Rev.D 67, 049902
  (2003)]},\ \Eprint {http://arxiv.org/abs/hep-ph/9911494}
  {arXiv:hep-ph/9911494} \BibitemShut {NoStop}%
\bibitem [{\citenamefont {Weldon}(2005)}]{PhysRevD.72.096005}%
  \BibitemOpen
  \bibfield  {author} {\bibinfo {author} {\bibfnamefont {H.~A.}\ \bibnamefont
  {Weldon}},\ }\href {\doibase 10.1103/PhysRevD.72.096005} {\bibfield
  {journal} {\bibinfo  {journal} {Phys. Rev. D}\ }\textbf {\bibinfo {volume}
  {72}},\ \bibinfo {pages} {096005} (\bibinfo {year} {2005})}\BibitemShut
  {NoStop}%
\bibitem [{\citenamefont {Bodeker}\ and\ \citenamefont
  {Sangel}(2017)}]{Bodeker:2017deo}%
  \BibitemOpen
  \bibfield  {author} {\bibinfo {author} {\bibfnamefont {D.}~\bibnamefont
  {Bodeker}}\ and\ \bibinfo {author} {\bibfnamefont {M.}~\bibnamefont
  {Sangel}},\ }\href {\doibase 10.1088/1475-7516/2017/06/052} {\bibfield
  {journal} {\bibinfo  {journal} {JCAP}\ }\textbf {\bibinfo {volume} {1706}},\
  \bibinfo {pages} {052} (\bibinfo {year} {2017})},\ \Eprint
  {http://arxiv.org/abs/1702.02155} {arXiv:1702.02155 [hep-ph]} \BibitemShut
  {NoStop}%
\bibitem [{\citenamefont {Sternbeck}\ \emph {et~al.}(2006)\citenamefont
  {Sternbeck}, \citenamefont {Ilgenfritz}, \citenamefont {Muller-Preussker},
  \citenamefont {Schiller},\ and\ \citenamefont
  {Bogolubsky}}]{Sternbeck:2006cg}%
  \BibitemOpen
  \bibfield  {author} {\bibinfo {author} {\bibfnamefont {A.}~\bibnamefont
  {Sternbeck}}, \bibinfo {author} {\bibfnamefont {E.~M.}\ \bibnamefont
  {Ilgenfritz}}, \bibinfo {author} {\bibfnamefont {M.}~\bibnamefont
  {Muller-Preussker}}, \bibinfo {author} {\bibfnamefont {A.}~\bibnamefont
  {Schiller}}, \ and\ \bibinfo {author} {\bibfnamefont {I.~L.}\ \bibnamefont
  {Bogolubsky}},\ }\href@noop {} {\bibfield  {journal} {\bibinfo  {journal}
  {PoS}\ }\textbf {\bibinfo {volume} {LAT2006}},\ \bibinfo {pages} {076}
  (\bibinfo {year} {2006})},\ \Eprint {http://arxiv.org/abs/hep-lat/0610053}
  {arXiv:hep-lat/0610053} \BibitemShut {NoStop}%
\bibitem [{\citenamefont {Gao}\ \emph {et~al.}(2021)\citenamefont {Gao},
  \citenamefont {Papavassiliou},\ and\ \citenamefont
  {Pawlowski}}]{Gao:2021wun}%
  \BibitemOpen
  \bibfield  {author} {\bibinfo {author} {\bibfnamefont {F.}~\bibnamefont
  {Gao}}, \bibinfo {author} {\bibfnamefont {J.}~\bibnamefont {Papavassiliou}},
  \ and\ \bibinfo {author} {\bibfnamefont {J.~M.}\ \bibnamefont {Pawlowski}},\
  }\href {\doibase 10.1103/PhysRevD.103.094013} {\bibfield  {journal} {\bibinfo
   {journal} {Phys. Rev. D}\ }\textbf {\bibinfo {volume} {103}},\ \bibinfo
  {pages} {094013} (\bibinfo {year} {2021})},\ \Eprint
  {http://arxiv.org/abs/2102.13053} {arXiv:2102.13053 [hep-ph]} \BibitemShut
  {NoStop}%
\bibitem [{\citenamefont {Falc\~ao}\ \emph {et~al.}(2020)\citenamefont
  {Falc\~ao}, \citenamefont {Oliveira},\ and\ \citenamefont
  {Silva}}]{Falcao:2020vyr}%
  \BibitemOpen
  \bibfield  {author} {\bibinfo {author} {\bibfnamefont {A.~F.}\ \bibnamefont
  {Falc\~ao}}, \bibinfo {author} {\bibfnamefont {O.}~\bibnamefont {Oliveira}},
  \ and\ \bibinfo {author} {\bibfnamefont {P.~J.}\ \bibnamefont {Silva}},\
  }\href {\doibase 10.1103/PhysRevD.102.114518} {\bibfield  {journal} {\bibinfo
   {journal} {Phys. Rev. D}\ }\textbf {\bibinfo {volume} {102}},\ \bibinfo
  {pages} {114518} (\bibinfo {year} {2020})},\ \Eprint
  {http://arxiv.org/abs/2008.02614} {arXiv:2008.02614 [hep-lat]} \BibitemShut
  {NoStop}%
\bibitem [{\citenamefont {Dudal}\ \emph {et~al.}(2011)\citenamefont {Dudal},
  \citenamefont {Guimaraes},\ and\ \citenamefont {Sorella}}]{Dudal:2010cd}%
  \BibitemOpen
  \bibfield  {author} {\bibinfo {author} {\bibfnamefont {D.}~\bibnamefont
  {Dudal}}, \bibinfo {author} {\bibfnamefont {M.}~\bibnamefont {Guimaraes}}, \
  and\ \bibinfo {author} {\bibfnamefont {S.}~\bibnamefont {Sorella}},\ }\href
  {\doibase 10.1103/PhysRevLett.106.062003} {\bibfield  {journal} {\bibinfo
  {journal} {Phys.Rev.Lett.}\ }\textbf {\bibinfo {volume} {106}},\ \bibinfo
  {pages} {062003} (\bibinfo {year} {2011})},\ \Eprint
  {http://arxiv.org/abs/1010.3638} {arXiv:1010.3638 [hep-th]} \BibitemShut
  {NoStop}%
\bibitem [{\citenamefont {Meyers}\ and\ \citenamefont
  {Swanson}(2013)}]{Meyers:2012ka}%
  \BibitemOpen
  \bibfield  {author} {\bibinfo {author} {\bibfnamefont {J.}~\bibnamefont
  {Meyers}}\ and\ \bibinfo {author} {\bibfnamefont {E.~S.}\ \bibnamefont
  {Swanson}},\ }\href {\doibase 10.1103/PhysRevD.87.036009} {\bibfield
  {journal} {\bibinfo  {journal} {Phys. Rev. D}\ }\textbf {\bibinfo {volume}
  {87}},\ \bibinfo {pages} {036009} (\bibinfo {year} {2013})},\ \Eprint
  {http://arxiv.org/abs/1211.4648} {arXiv:1211.4648 [hep-ph]} \BibitemShut
  {NoStop}%
\bibitem [{\citenamefont {Meyer}\ and\ \citenamefont
  {Swanson}(2015)}]{Meyer:2015eta}%
  \BibitemOpen
  \bibfield  {author} {\bibinfo {author} {\bibfnamefont {C.~A.}\ \bibnamefont
  {Meyer}}\ and\ \bibinfo {author} {\bibfnamefont {E.~S.}\ \bibnamefont
  {Swanson}},\ }\href {\doibase 10.1016/j.ppnp.2015.03.001} {\bibfield
  {journal} {\bibinfo  {journal} {Prog. Part. Nucl. Phys.}\ }\textbf {\bibinfo
  {volume} {82}},\ \bibinfo {pages} {21} (\bibinfo {year} {2015})},\ \Eprint
  {http://arxiv.org/abs/1502.07276} {arXiv:1502.07276 [hep-ph]} \BibitemShut
  {NoStop}%
\bibitem [{\citenamefont {Sanchis-Alepuz}\ \emph {et~al.}(2015)\citenamefont
  {Sanchis-Alepuz}, \citenamefont {Fischer}, \citenamefont {Kellermann},\ and\
  \citenamefont {von Smekal}}]{Sanchis-Alepuz:2015hma}%
  \BibitemOpen
  \bibfield  {author} {\bibinfo {author} {\bibfnamefont {H.}~\bibnamefont
  {Sanchis-Alepuz}}, \bibinfo {author} {\bibfnamefont {C.~S.}\ \bibnamefont
  {Fischer}}, \bibinfo {author} {\bibfnamefont {C.}~\bibnamefont {Kellermann}},
  \ and\ \bibinfo {author} {\bibfnamefont {L.}~\bibnamefont {von Smekal}},\
  }\href {\doibase 10.1103/PhysRevD.92.034001} {\bibfield  {journal} {\bibinfo
  {journal} {Phys. Rev. D}\ }\textbf {\bibinfo {volume} {92}},\ \bibinfo
  {pages} {034001} (\bibinfo {year} {2015})},\ \Eprint
  {http://arxiv.org/abs/1503.06051} {arXiv:1503.06051 [hep-ph]} \BibitemShut
  {NoStop}%
\bibitem [{\citenamefont {Souza}\ \emph {et~al.}(2020)\citenamefont {Souza},
  \citenamefont {Narciso~Ferreira}, \citenamefont {Aguilar}, \citenamefont
  {Papavassiliou}, \citenamefont {Roberts},\ and\ \citenamefont
  {Xu}}]{Souza:2019ylx}%
  \BibitemOpen
  \bibfield  {author} {\bibinfo {author} {\bibfnamefont {E.~V.}\ \bibnamefont
  {Souza}}, \bibinfo {author} {\bibfnamefont {M.}~\bibnamefont
  {Narciso~Ferreira}}, \bibinfo {author} {\bibfnamefont {A.~C.}\ \bibnamefont
  {Aguilar}}, \bibinfo {author} {\bibfnamefont {J.}~\bibnamefont
  {Papavassiliou}}, \bibinfo {author} {\bibfnamefont {C.~D.}\ \bibnamefont
  {Roberts}}, \ and\ \bibinfo {author} {\bibfnamefont {S.-S.}\ \bibnamefont
  {Xu}},\ }\href {\doibase 10.1140/epja/s10050-020-00041-y} {\bibfield
  {journal} {\bibinfo  {journal} {Eur. Phys. J. A}\ }\textbf {\bibinfo {volume}
  {56}},\ \bibinfo {pages} {25} (\bibinfo {year} {2020})},\ \Eprint
  {http://arxiv.org/abs/1909.05875} {arXiv:1909.05875 [nucl-th]} \BibitemShut
  {NoStop}%
\bibitem [{\citenamefont {Huber}\ \emph {et~al.}(2020)\citenamefont {Huber},
  \citenamefont {Fischer},\ and\ \citenamefont
  {Sanchis-Alepuz}}]{Huber:2020ngt}%
  \BibitemOpen
  \bibfield  {author} {\bibinfo {author} {\bibfnamefont {M.~Q.}\ \bibnamefont
  {Huber}}, \bibinfo {author} {\bibfnamefont {C.~S.}\ \bibnamefont {Fischer}},
  \ and\ \bibinfo {author} {\bibfnamefont {H.}~\bibnamefont {Sanchis-Alepuz}},\
  }\href {\doibase 10.1140/epjc/s10052-020-08649-6} {\bibfield  {journal}
  {\bibinfo  {journal} {Eur. Phys. J. C}\ }\textbf {\bibinfo {volume} {80}},\
  \bibinfo {pages} {1077} (\bibinfo {year} {2020})},\ \Eprint
  {http://arxiv.org/abs/2004.00415} {arXiv:2004.00415 [hep-ph]} \BibitemShut
  {NoStop}%
\bibitem [{\citenamefont {Athenodorou}\ and\ \citenamefont
  {Teper}(2020)}]{Athenodorou:2020ani}%
  \BibitemOpen
  \bibfield  {author} {\bibinfo {author} {\bibfnamefont {A.}~\bibnamefont
  {Athenodorou}}\ and\ \bibinfo {author} {\bibfnamefont {M.}~\bibnamefont
  {Teper}},\ }\href {\doibase 10.1007/JHEP11(2020)172} {\bibfield  {journal}
  {\bibinfo  {journal} {JHEP}\ }\textbf {\bibinfo {volume} {11}},\ \bibinfo
  {pages} {172} (\bibinfo {year} {2020})},\ \Eprint
  {http://arxiv.org/abs/2007.06422} {arXiv:2007.06422 [hep-lat]} \BibitemShut
  {NoStop}%
\bibitem [{\citenamefont {Sanchis-Alepuz}\ and\ \citenamefont
  {Williams}(2015)}]{Sanchis-Alepuz:2015tha}%
  \BibitemOpen
  \bibfield  {author} {\bibinfo {author} {\bibfnamefont {H.}~\bibnamefont
  {Sanchis-Alepuz}}\ and\ \bibinfo {author} {\bibfnamefont {R.}~\bibnamefont
  {Williams}},\ }\href {\doibase 10.1088/1742-6596/631/1/012064} {\bibfield
  {journal} {\bibinfo  {journal} {J. Phys. Conf. Ser.}\ }\textbf {\bibinfo
  {volume} {631}},\ \bibinfo {pages} {012064} (\bibinfo {year} {2015})},\
  \Eprint {http://arxiv.org/abs/1503.05896} {arXiv:1503.05896 [hep-ph]}
  \BibitemShut {NoStop}%
\bibitem [{\citenamefont {Horak}\ \emph {et~al.}()\citenamefont {Horak},
  \citenamefont {Pawlowski},\ and\ \citenamefont {Wink}}]{horak:preparation}%
  \BibitemOpen
  \bibfield  {author} {\bibinfo {author} {\bibfnamefont {J.}~\bibnamefont
  {Horak}}, \bibinfo {author} {\bibfnamefont {J.~M.}\ \bibnamefont
  {Pawlowski}}, \ and\ \bibinfo {author} {\bibfnamefont {N.}~\bibnamefont
  {Wink}},\ }\href@noop {} {\ }\bibinfo {note} {\textit{in
  preparation}}\BibitemShut {NoStop}%
\end{thebibliography}
%

\end{document}